\documentclass[final,onecolumn,prfluids,aps,amsfonts,amssymb,showpacs,floatfix,nofootinbib]{revtex4-2}
\usepackage{amsmath}
\usepackage[integrals]{wasysym}
\usepackage{graphicx}
\usepackage[dvipsnames]{color}
\usepackage{hyperref}
\usepackage{psfrag}
\usepackage{stmaryrd}
\usepackage{amssymb}
\usepackage{float}
\usepackage{color}
\usepackage{comment}
\usepackage[T1]{fontenc}
\usepackage[french,english]{babel}
\usepackage[utf8]{inputenc}
\usepackage{lmodern}
\usepackage{multirow, makecell}

\usepackage{geometry}
\begin{document}

\title{Internal wave turbulence in a stratified fluid with and without eigenmodes of the experimental domain}

\author{Nicolas Lanchon}
\affiliation{Universit\'e Paris-Saclay, CNRS, FAST, 91405 Orsay, France}
\author{Daniel Odens Mora}
\affiliation{Universit\'e Paris-Saclay, CNRS, FAST, 91405 Orsay, France}
\author{Eduardo Monsalve}
\affiliation{Universit\'e Paris-Saclay, CNRS, FAST, 91405 Orsay, France}
\author{Pierre-Philippe~Cortet}
\email[]{pierre-philippe.cortet@universite-paris-saclay.fr}
\affiliation{Universit\'e Paris-Saclay, CNRS, FAST, 91405 Orsay, France}

\date{\today}

\begin{abstract}
We present laboratory experiments on turbulence in a linearly stratified fluid driven by an ensemble of internal gravity waves which approaches statistical homogeneity and axi-symmetry. In a way similar to several recent experimental works, non-linearities develop through the establishment of a set of internal wave modes at discrete frequencies, when the forcing amplitude is increased. We show that the most energetic of these modes are resonant eigenmodes of the fluid domain. The discretization of the energy in frequency and wavenumber associated to the emergence of these modes prevents the flow from approaching a regime described by the Weak/Wave Turbulence Theory, in which a forward cascade carried by a statistical ensemble of weakly non-linear waves in an infinite domain forms an energy continuum in the frequency and wavenumber spaces. We then show that the introduction of slightly tilted panels at the top and at the bottom of the fluid domain allows to inhibit the emergence of the discrete wave modes. In this new configuration, the non-linear regime results in a continuum of energy over one decade of frequencies which is mainly carried by internal gravity waves verifying the dispersion relation. We therefore achieved a turbulent flow approaching a three-dimensional internal wave turbulence regime with no discretization of the energy in the frequency and wavenumber domains. These results constitute a significant step forward in the search of the laboratory observation of a fully-developed weakly-non-linear internal-gravity-wave turbulence.
\end{abstract}

\maketitle

\section{Introduction}

A stable stratification of the fluid density deeply modifies the features of hydrodynamic turbulence~\cite{Davidson2013,Riley2012}, which in the first place becomes anisotropic. A major change is the fact that stratification allows the propagation of waves in the bulk of the fluid~\cite{Staquet2002,Sutherland2010,Dauxois2018}. These waves are called internal gravity waves and result from the restoring action of the buoyancy force. They are dispersive and anisotropic and possess peculiar features: a wavelength independent of the wave frequency, group and phase velocities normal to each other, and a direction of propagation set by the wave frequency. In the model case of a linearly stratified fluid, the inviscid dispersion relation writes~\cite{Staquet2002,Sutherland2010,Dauxois2018} 
\begin{equation}\label{eq:disp}
    \sigma= \pm N \frac{k_\perp}{\sqrt{k_\perp^2+k_\parallel^2}}
\end{equation}
where $\sigma$ is the wave angular frequency, and $k_\perp$ and $k_\parallel$ are the norm of the components of the wavevector ${\bf k}$ normal and parallel to gravity, respectively. The buoyancy frequency $N=\sqrt{-g/\rho_0\,d\rho/dz}$ is set by the intensity of the density gradient at rest, $d\rho/dz<0$, and the acceleration of gravity $g$ (with the vertical coordinate $z$ oriented opposite to gravity). Equation~(\ref{eq:disp}) is obtained from Navier-Stokes equations under the Boussinesq approximation which consists in restricting to weak density variations with respect to the reference density $\rho_0$~\cite{Sutherland2010,Dauxois2018}.

Together with the Earth rotation, the stratification in density of geophysical fluids is a key and ubiquitous ingredient of the oceanic and atmospheric turbulent dynamics~\cite{Pedlovsky1987,Vallis2006,Wunsch2004,MacKinnon2017}. 
In this context, a major feature of global oceanic and atmospheric models is the use of parametrizations in order to account for the ``small'' scales~\cite{Stensrud2007,Polzin2014,MacKinnon2017,Gregg2018}, at which the fluid stratification is in general expected to play a major role. Progress in the modeling of these ``small-scale'' dynamics constitutes a long-standing lever to improve oceanic and atmospheric simulations, which has motivated the search for a fundamental understanding of ``stratified turbulence''.

Turbulence in a stably stratified fluid can develop in several different regimes, which can coexist over different ranges of scales. This complexity is notably related to the fact that stratified fluids allow the propagation of internal waves, which can coexist and interact with vortex structures. In order to anticipate which regime will be at play, one should consider three independent non-dimensional numbers: the classical Reynolds number $Re= u \ell/\nu$, the Froude number $Fr=u/N \ell$ and the non-dimensional frequency $\sigma^*=\sigma/N$, where $u$ and $\sigma$ are the characteristic velocity and evolution rate of the flow structures at scale $\ell$ and $\nu$ is the fluid kinematic viscosity. Moreover, since stratified turbulence is anisotropic, several variants of these non-dimensional numbers can be relevant in order to properly account for the different roles of the horizontal $u_\perp$ and vertical $u_\parallel$ velocity components and of the horizontal $\ell_\perp$ and vertical $\ell_\parallel$ scales.

From a fundamental point of view, a regime which has received a lot of attention is the so-called ``strongly stratified turbulence'' (SST)~\cite{Brethouwer2007,Davidson2013}. In this regime, the flow is dominated by horizontally elongated ``pancake'' eddies (with $\ell_\parallel\ll\ell_\perp$ and $u_\parallel\ll u_\perp$) which are shearing each other in the vertical direction. Phenomenological predictions for this regime, expected to produce a direct cascade of energy from large to small scales, have been put forward for the 1D spatial kinetic energy spectra~\cite{Dewan1997,Lindborg2006,Brethouwer2007,Nazarenko2011a}: A Kolmogorov-like scaling for the spectrum as a function of the horizontal wavenumber, $E(k_\perp) \sim \epsilon^{2/3} k_\perp^{-5/3}$, where $\epsilon$ is the mean kinetic energy dissipation rate (per unit mass), and a scaling involving the buoyancy frequency $N$ for the spectrum as a function of the vertical wavenumber, $E(k_\parallel) \sim N^2 k_\parallel^{-3}$. This vertical spectrum is often referred to as the ``saturation spectrum''~\cite{Waite2006,Rorai2015} as a consequence of an early justification for it based on the concept of ``saturated'' internal gravity waves at the onset of breaking~\cite{Dewan1997}. Nevertheless, a more global justification of the whole SST phenomenology on the basis of a self-similarity of the system of dynamical equations has later been proposed during the 2000's~\cite{Billant2001,Lindborg2006,Brethouwer2007} and can also be found from the ``critical balance'' phenomenology~\cite{Nazarenko2011a}.
The SST regime is expected for large ``horizontal'' Reynolds numbers $Re_\perp=u_\perp \ell_\perp/\nu \gg 1$ and low ``horizontal'' Froude numbers $Fr_\perp =u_\perp/N \ell_\perp \ll 1$ provided their combination $Re_b=Re_\perp Fr^2_\perp$, often called the buoyancy Reynolds number, is large~\cite{Brethouwer2007}. Besides, it is expected that the vertical integral scale of the turbulence, i.e. the vertical thickness of the ``pancake eddies'', dynamically adjusts to the so-called buoyancy scale $u_\perp/N$~\cite{Billant2001,Riley2003,Brethouwer2007,Davidson2013}, which process can for instance result from the zig-zag instability discovered by Billant and Chomaz~\cite{Billant2000a,Billant2000b}. This vertical scale selection eventually leads the ``vertical'' Froude number $Fr_\parallel=u_\perp/N\ell_\parallel$ to be of order $1$ at the integral scale, this result being also true on a scale-by-scale basis. The predictions for the spatial kinetic energy spectra in this strongly stratified regime are consistent with the results of several direct numerical simulations~\cite{Lindborg2006,Brethouwer2007,Rorai2015,Augier2015,Maffioli2017}. Besides, the predictions of the SST phenomenology are also remarkable because they seem to explain observations at small and mesoscales in the atmosphere~\cite{Nastrom1985,Dewan1986,Dewan1997,Cot2001,Brethouwer2007,Riley2008} as well as at small scales in the oceans (see Ref.~\cite{Riley2008} and references therein).

In the SST phenomenology, the rate of evolution $\sigma$ of the turbulent structures at scale $\ell_\perp$ is driven by the non-linear frequency $u_\perp/\ell_\perp$~\cite{Billant2001,Davidson2013} such that the turbulence is strongly non-linear, even if the horizontal Froude number $Fr_\perp = u_\perp/N \ell_\perp$ is very small. In the following, we still consider flows with large Reynolds numbers and small Froude numbers at the injection scale. However, the flows will be weakly non-linear: This implies a separation between the linear timescale $1/\sigma$ and the non-linear timescale $\ell/u\gg 1/\sigma$ which is a characteristic of ``Wave Turbulence''~\cite{Nazarenko2011} (we simply use here the typical velocity $u$ of the waves of wavelength $\ell$ noticing that, for a plane internal wave, we have the relations $u_\perp/u_\parallel=k_\parallel/k_\perp=({\sigma^*}^{-2}-1)^{1/2}$ with $0\leq \sigma^*\leq 1$). This timescale separation is equivalent to having a Froude number $Fr=u/N \ell$ much smaller than the non-dimensional frequency $\sigma^*=\sigma/N$ which is itself bounded by $1$ for internal waves. In this situation, another fundamental regime of stratified turbulence is expected: the ``Wave Turbulence'' regime~\cite{Nazarenko2011}. In this weakly non-linear regime, the energy cascade is carried by a statistical ensemble of weakly non-linear internal gravity waves. The energy is transferred, on statistical average, from large to small scales at a rate much lower than the wave frequencies. Because of the quadratic non-linearity of the Navier-Stokes equation, this forward energy cascade is expected to result from triadic resonant interactions between internal waves~\cite{Lvov2010,Nazarenko2011}.

As a consequence, in parallel to the SST regime, theoretical research has emerged and is still active on the wave turbulence regime in stratified fluids (also called weak turbulence)~\cite{Pelinovsky1977,McComas1981,Caillol2000,Lvov2001,Lvov2004,Lvov2010,Dematteis2021}. An underlying motivation is the understanding of the oceanic dynamics and energy spectra at small scales, whose most famous empirical description is the Garrett-and-Munk small-scale high-frequency spectrum $E(k_\parallel,\sigma)\sim k_\parallel^{-2}\sigma^{-2}$~\cite{Garrett1979,Polzin2011}. This spectrum is thought to result from the non-linear dynamics of internal gravity waves~\cite{Polzin2011} producing a forward energy cascade (which might eventually feed the above-mentioned SST cascade of energy observed at even smaller oceanic scales~\cite{Riley2008}). The Wave Turbulence Theory (WTT) is appealing because of its analytical nature and because it has already been successfully applied to some other wave systems, such as capillary surface waves~\cite{Falcon2022} and inertial waves in rotating fluids~\cite{Monsalve2020}. Theoretically, we still expect, in the wave turbulence regime, a direct and anisotropic energy cascade from large to small scales but with different scaling laws (than for SST) for the rate of energy transfer and the energy spectra~\cite{Caillol2000,Nazarenko2011}. Following the canonical derivation of the wave turbulence formalism, one predicted in the limit $k_\perp \ll k_\parallel$ a constant flux solution with a 2D spatial kinetic energy spectrum scaling as $E(k_\perp,k_\parallel) \sim \sqrt{N \epsilon}\,k_\perp^{-3/2} k_\parallel^{-3/2}$~\cite{Caillol2000,Lvov2001}. Nevertheless, the implementation of the WTT in the case of stratified fluids later revealed to be more complex and is still the subject of delicate analytical questions regarding the convergence of the collision integral~\cite{Dematteis2021}. In practice, more advanced theoretical works suggested a family of solutions for the 2D spatial kinetic energy spectrum~\cite{Lvov2004,Lvov2010} among which the scaling law $E(k_\perp,k_\parallel) \sim k_\perp^{2-a} k_\parallel^{-1}$ with $a\simeq 3.69$ would seem to be the option to prefer considering convergence conditions for the collision integral~\cite{Lvov2010,Dematteis2021}. Remarkably, this last prediction (with the exponent $a\simeq 3.69$) is relatively close to the Garrett-and-Munk spectrum (which corresponds to the case $a=4$). In this complex theoretical framework, experiments and numerical simulations promoting the emergence of a turbulent flow in the regime of internal wave turbulence are of strong interest to guide future theoretical developments and improve our understanding of the oceanic dynamics.

Stratified turbulence obtained from energy injection in weakly non-linear internal gravity waves has nevertheless only recently been achieved experimentally and numerically. This is possibly due to the fact it implies large facilities in the experimental case (in order to simultaneously reach large $Re$ and small $Fr$) and important computational resources in the numerical case (notably because of the linear/non-linear timescale separation). For instance, Le Reun~\textit{et al.}~\cite{LeReun2018} studied via direct numerical simulations of the Boussinesq equations a turbulent state in which the energy is indirectly injected in weakly non-linear internal waves through a parametric instability driven by a periodic tidal deformation. Le Reun and coworkers report a scenario different from the one observed in the SST regime in previous simulations~\cite{Lindborg2006,Brethouwer2007,Rorai2015,Maffioli2017}: they observe an energy cascade carried almost only by internal waves, i.e. structures verifying the wave dispersion relation. They report 1D spatial energy spectra with scaling laws in $k_\perp^{-3}$ and $k_\parallel^{-3}$, and a dominant transfer of energy toward frequencies smaller than the forcing frequency $\sigma_0^*$. In addition to these features, a weakly-energetic power-law behavior of the temporal energy spectrum, compatible with an exponent $-2$, is observed at frequencies larger than the forcing frequency $\sigma_0^*$. This power law, which is observed only when $\sigma_0^*$ is significantly smaller than $1$ (the limit frequency for internal waves) could be the sign of a dynamics similar to that behind the oceanic Garrett-and-Munk celebrated spectrum~\cite{Polzin2011}.

On the experimental side, Savaro \textit{et al.}~\cite{Savaro2020} recently conducted experiments in a stratified fluid, where energy is injected in large-scale internal waves in the frequency range $0.65 \leq \sigma_0^*\leq 0.75$ at large Reynolds number, low Froude number and buoyancy Reynolds number ranging from $1$ to $30$. At moderate forcing amplitude, the authors report the emergence of a series of internal wave eigenmodes of the fluid domain at subharmonic frequencies (below $\sigma_0^*$). Then, as the forcing amplitude is increased, a continuum of energy at subharmonic frequencies, compatible with an ensemble of propagating internal waves, develops and tends to progressively take over the eigenmodes in the energy budget. As this transition proceeds, eddies and mixing events seem to emerge at small scales reducing the extension of the energy cascade toward small scales and suggesting that the turbulent flow is close to transition to a strongly non-linear regime.

In a subsequent paper, Rodda \textit{et al.}~\cite{Rodda2022} report experiments in a modified version of the setup of Savaro \textit{et al.}~\cite{Savaro2020} where the shape of the water tank has been changed from square to pentagonal. A second change is that the forcing injects now energy in large-scale internal waves at a lower non-dimensional frequency $\sigma_0^*$, in the range $0.16\leq \sigma_0^*\leq 0.38$. These experiments led, in the non-linear regime, to a temporal kinetic energy spectrum dominated by a series of peaks at harmonic frequencies of the forcing frequency $\sigma_0^*$. Remarkably, when the forcing Froude number is further increased, this discrete spectrum superimposes with a continuum of energy at frequencies larger than the forcing frequency $\sigma_0^*$ which is compatible with a power law $\sigma^{-2}$ and therefore with the Garrett-and-Munk spectrum. This emerging power-law behavior, which also reminds the observations of Le Reun \textit{et al.}~\cite{LeReun2018}, extends here beyond the limit frequency of internal waves $\sigma=N$ whereas Le Reun~\textit{et al.} observed a clear cut-off at $\sigma=N$ in their spectrum. According to Rodda \textit{et al.}, this extension of the $\sigma^{-2}$ power law above the buoyancy frequency $N$ might indicate that the flow is evolving toward a strongly nonlinear turbulence.

Another remarkable recent experimental contribution is the internal wave attractor experiment by Davis \textit{et al.}~\cite{Davis2020}. Davis and coworkers use a wave maker to force a large-scale internal gravity wave at the specific frequency $\sigma_0^* = 0.62$. The experiments are conducted in a trapezoidal cavity which is nearly 2D, i.e. thin in the horizontal direction normal to the vertical trapeze. Due to the unusual reflection laws of internal waves, this shape leads the forced wave to focus on a ``wave attractor''~\cite{Maas1995,Maas1997,Grisouard2008}. This base flow consists in a mono-frequency self-similar internal wave beam resulting from an equilibrium between viscous dissipation and wave focusing at reflection on the tilted surface~\cite{Grisouard2008,Hazewinkel2008,Brunet2019}. As the forcing amplitude is increased, non-linear effects emerge first through a classical triadic resonance instability of the primary wave beam producing new internal waves at two subharmonic frequencies~\cite{Dauxois2018}. Then, secondary triadic interactions lead to the emergence, in the temporal energy spectrum, of a series of discrete energy peaks at subharmonic frequencies (see~\cite{Davis2019} for details) in a way similar to Savaro \textit{et al.}~\cite{Savaro2020}. At large forcing amplitude, Davis \textit{et al.}~\cite{Davis2020} report a behavior compatible with $k^{-3}$ for the 1D spatial kinetic energy spectrum averaged over all directions (in the vertical plane), which result is compatible with the observations of Le Reun~\textit{et al.}~\cite{LeReun2018}.

To conclude, although these experimental and numerical works tend to approach an internal gravity wave turbulence regime, it is still uncertain whether the observed flows are consistent with the regime described by the WTT framework. This gap might be due to a still too strong influence of finite size effects, viscous effects and/or strong non-linearities in the experiments and the numerics.

In this article, we report results obtained with an experimental setup designed to generate a weakly non-linear turbulence regime in a linearly stratified fluid. Our forcing device injects energy in an ensemble of internal gravity waves at frequency $\sigma_0^*=0.94$ whose statistics tends to be homogeneous and axisymmetric (around the vertical), two assumptions made when deriving WTT~\cite{Caillol2000,Lvov2004,Lvov2010,Dematteis2021}. This forcing device has recently allowed an observation in the lab of an inertial wave turbulence regime~\cite{Monsalve2020} matching quantitatively the theoretical predictions of WTT for rotating fluids~\cite{Galtier2003}.

In a first series of experiments, we report, similarly to Savaro \textit{et al.}~\cite{Savaro2020} and Davis \textit{et al.}~\cite{Davis2020}, the emergence in the nonlinear regime of a set of subharmonic internal wave modes at discrete subharmonic frequencies. The most energetic of these modes reveal to be internal wave eigenmodes of the fluid domain following from the finite size of our system. By discretizing the energy in the frequency and wavenumber domain, this feature prevents any comparison with WTT which is built under the assumption of an ensemble of random and propagative waves in an infinite domain which forms an energy continuum in the frequency and wavenumber spaces. Nevertheless, we then present a slight but crucial modification in the shape of our fluid domain which succeed in preventing the emergence of the fluid domain eigenmodes. In this new configuration, the non-linear regime gives birth to a continuum of energy over typically one decade of subharmonic frequencies. We show that this energy continuum is carried by structures at scales smaller than the forced wave mode and verifying the internal wave dispersion relation. The flow observed at the largest forcing Reynolds number reveals horizontal and vertical 1D spatial kinetic energy spectra compatible with a $k^{-3}$ behavior.

\section{Experimental setup}\label{sec:setup}

\begin{figure}
	\centerline{\includegraphics[width=10cm]{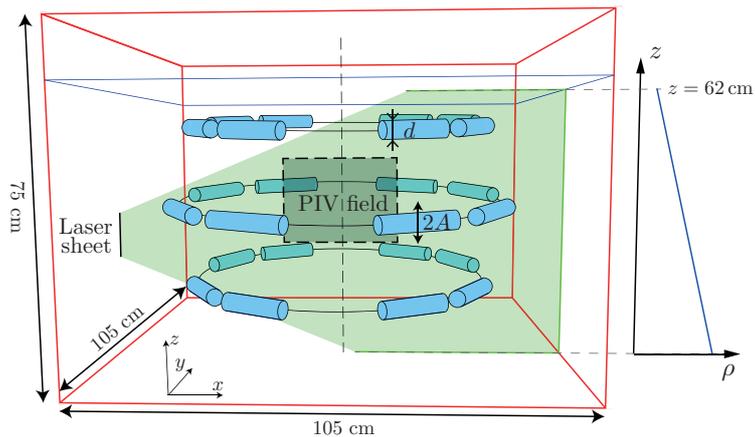}}
	\caption{Experimental setup. 24 horizontal cylinders oscillate vertically, i.e. along the $z$-axis, inside a glass tank filled up to a height of $62$~cm with a linearly stably stratified fluid of buoyancy frequency $N \simeq 1$~rad/s.}\label{fig_expset}
\end{figure}

The experimental setup, sketched in Fig.~\ref{fig_expset}, consists of twenty-four horizontal cylinders oscillating vertically inside a glass tank of $105 \times 105$~cm$^2$ base filled up to a height of $62$~cm with a linearly stratified fluid of buoyancy frequency $N$ close to $1$~rad/s. This forcing device is similar to that of the rotating fluid experiments reported in Refs.~\cite{Brunet2020,Monsalve2020}. The cylinders, which have a diameter $d=4$~cm, are evenly organized on three ``parallels'' of an $80$-cm-diameter virtual sphere centered in the tank. Each parallel contains eight cylinders which are arranged as follows: $18$-cm-long cylinders are regularly distributed on the virtual sphere equator at a height of $36$~cm above the bottom of the tank, and $15$-cm-long cylinders are regularly distributed on the two parallels placed at a vertical distance of $19$~cm above and below the virtual sphere equator. Each cylinder is actuated in a sinusoidal vertical oscillating translation of amplitude $A$ and angular frequency $\sigma_0=0.94 \times N$, with a random initial phase set independently for each cylinder. In the linear regime, at small forcing amplitude $A$, an oscillating cylinder produces self-similar internal gravity wave beams, which propagate in the four directions normal to the cylinder axis and which make an angle $\cos^{-1}(\sigma_0/N)\simeq 20 ^\circ$ with respect to the vertical~\cite{Mowbray1967, Thomas1972, Sutherland2002, Gostiaux2007, Ermanyuk2008}. Taken all together, the geometrical arrangement of our forcing device aims to produce, in the central region of the experiment, a flow made of an ensemble of internal gravity waves which approaches statistical homogeneity and axi-symmetry around the vertical axis.

We fabricate the linearly stably stratified fluid by means of the classical \textit{double-bucket} method~\cite{Oster1963, Hill2002}: The first bucket is filled with a NaCl-water mixture whereas the second bucket is filled with water. Using this method, we obtain an experimental buoyancy frequency $N=\sqrt{-g/\rho_0 \, d \rho/d z}$ close to $1$~rad/s, where $\rho_0$ is the average density of the stratified fluid and $g$ the gravitational acceleration. To measure the vertical density profile of the fluid, we employ a Mettler Toledo conductivity probe (InLab 731-ISM-2m) mounted on a motorized translation stage. The probe is translated vertically at a velocity of $20$~mm/min from the tank bottom to the fluid free surface. During this translation, we record the conductivity and temperature that we convert into a measure of the fluid density using a calibration done with a densimeter (Anton Paar DMA35).

\begin{figure}
	\centerline{\includegraphics[width=14cm]{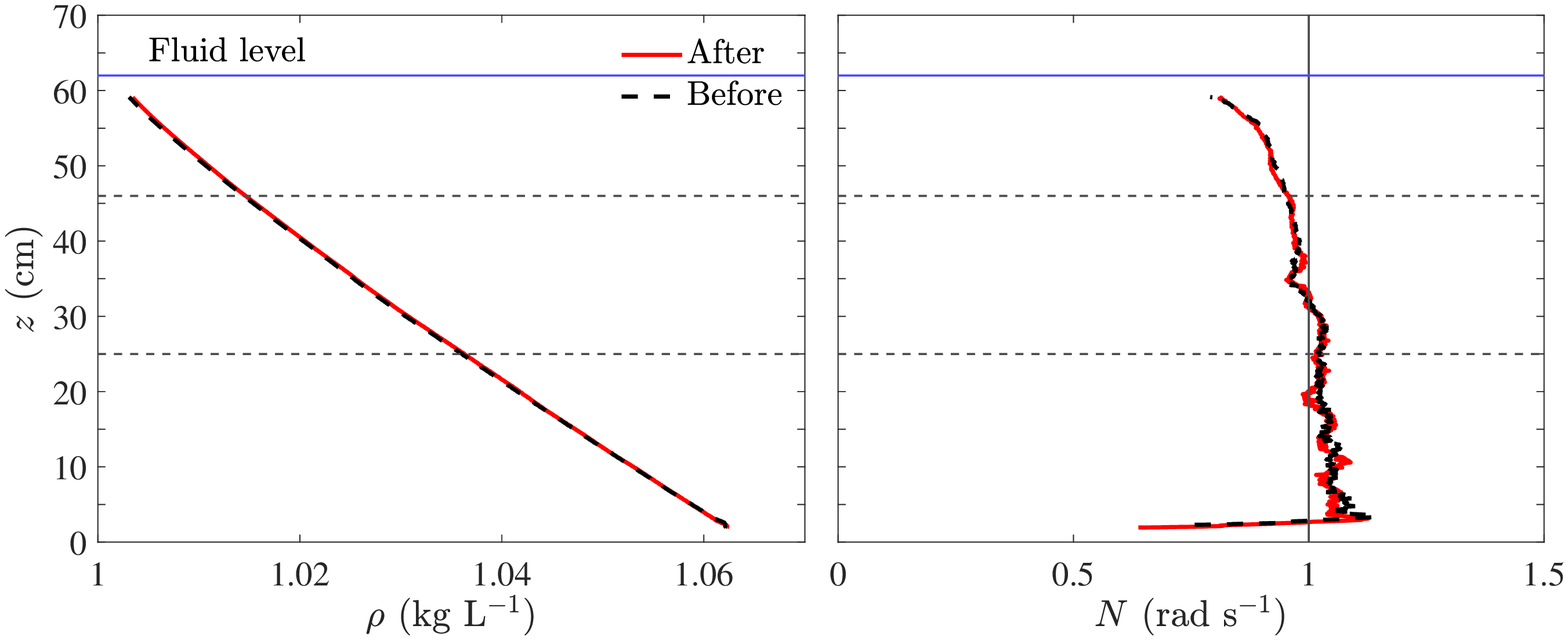}}
	\caption{Left panel: Vertical profile of the density $\rho$ before (black line) and after (red line) an experiment conducted with a forcing amplitude $A=2$~mm and which lasted $300$~periods of the forcing device.
	The vertical coordinate is noted $z$ and $z=0$ corresponds to the bottom of the experimental tank. The horizontal dashed lines show the vertical limits of the region imaged by the camera. Right panel: Corresponding vertical profiles of the buoyancy frequency $N=\sqrt{-g/\rho_0 \, d \rho/d z}$.}\label{fig_density}
\end{figure}

Figure~\ref{fig_density} shows the vertical profile of the density $\rho$ (left) and of the corresponding buoyancy frequency $N$ (right) before and after an experimental run of $300$~periods of the forcing $T=2\pi/\sigma_0$ at forcing amplitude $A=2$~mm. Figure~\ref{fig_density} reveals that both density profiles $\rho(z)$ are close to a linear behavior except for two layers of typical thickness of $2$ to $3$~cm at the top and at the bottom of the tank. In the region where the velocity measurements are taken, which has a vertical extent of $22$~cm (identified by the two horizontal dashed lines in Fig.~\ref{fig_density}), the buoyancy frequency $N$ is nearly uniform, slowly decreasing from $1.03$ to $0.96$~rad/s with increasing height $z$. In Fig.~\ref{fig_density}, it is clear that the density and buoyancy frequency profiles before and after the experiment are roughly identical which reveals that no significant mixing is induced by the flow. As we increase the forcing amplitude $A$, we observe that the density and buoyancy frequency profiles after the experiments start to slightly deviate from their initial profiles. At the largest explored forcing amplitude $A=12$~mm, we observe maximum local variations of the static density profile of about $0.5\permil$. These variations correspond to relative changes $\Delta N/N=~2\sqrt{\langle (N_{\rm after} - N_{\rm before})^2 \rangle_z}/{\langle N_{\rm after} + N_{\rm before} \rangle_z}$ of the buoyancy frequency ranging from $1\%$ at $A=2$~mm to $4\%$ at $A=12$~mm ($\langle \, \rangle_z$ stands for the spatial average over the vertical extent of the velocity measurement region).

During the experiments, the two components $u_x$ and $u_z$ of the velocity field are measured in a vertical plane at the center of the virtual sphere using a particle image velocimetry (PIV) system (see the PIV field in Fig.~\ref{fig_expset}). The fluid is seeded with 10-$\mu$m tracer particles and illuminated by a 140-mJ~Nd:YAG pulsed laser. Successive images covering an area of $\Delta x \times \Delta z= 289 \times 218$~mm$^2$ are recorded by a $2360 \times 1776$~pixels camera. PIV cross-correlation between successive images is applied using $32 \times 32$~pixels interrogation windows with $50\%$~overlap, providing velocity fields with a spatial resolution of $1.96$~mm. The image acquisition rate is adjusted, depending on the flow typical velocity, from $24$~images per forcing period at $A=2$~mm to $84$ images at $A=12$~mm. The PIV acquisitions start $30$ forcing periods $T=2\pi/\sigma_0$ before the forcing device is triggered, and then last between $330\,T$ and $830\,T$ depending on the experimental run.

\begin{table}
\caption{Parameters of the experiments reported in Sec.~\ref{sc:standingmodes}. $A$ is the cylinders amplitude of motion. $Re_f=A\sigma_0 d/\nu$ and $Fr_f=A\sigma_0/Nd$ are the forcing Reynolds and Froude numbers, $Re_{\rm rms}=u_{\rm rms}\lambda_0/\nu$, $Fr_{\rm rms}=u_{\rm rms}/N\lambda_0$ and $Re_b=Re_{\rm rms} Fr_{\rm rms}^2$, the Reynolds, Froude and buoyancy Reynolds numbers based on the flow rms velocity $u_{\rm rms}$ and $\lambda_0=15$~cm, the typical wavelength of the flow mode at the forcing frequency $\sigma_0=0.94\,N$. $d=4$~cm is the cylinders diameter, $\nu=10^{-6}$~m$^2$/s the average kinematic viscosity of the fluid and $N=1$~rad/s the average buoyancy frequency.}
\label{tab:freqb}
\begin{ruledtabular}
\begin{tabular}{ c c c c c c c c}
  $A$ (mm)   &  $Re_f$ & $Fr_f$  & $u_{\rm rms}$ (mm/s) & $Re_{\rm rms}$ & $Fr_{\rm rms}$ & $Re_b$ \\
\hline
 2  & 75 & 0.05 & 0.7 & 105 & 0.005 & 0.002 \\
 4 & 150 & 0.09 & 1.6 & 240 & 0.011 & 0.027 \\
 8 & 301 & 0.19 & 2.7 & 405 & 0.018 & 0.131 \\
 12 & 451 & 0.28 & 4.1 & 615 & 0.027 & 0.459 \\
\end{tabular}
\end{ruledtabular}
\end{table}

The parameters of the experiments reported in the next section 
(Sec.~\ref{sc:standingmodes}) are shown in Tab.~\ref{tab:freqb}. We define the 
forcing Froude number by $Fr_f=A\sigma_0/N d$ and the forcing Reynolds number 
by $Re_f=A\sigma_0d/\nu$, where $A$ is the cylinders amplitude of motion, $d$ 
the cylinders diameter, $N$ the buoyancy frequency, $\sigma_0=0.94\,N$ the 
forcing angular frequency and $\nu=10^{-6}$~m$^2$/s the average kinematic 
viscosity of the fluid. In Tab.~\ref{tab:freqb}, we also report the flow 
Reynolds number $Re_{\rm rms}=u_{\rm rms}\lambda_0/\nu$ and Froude number 
$Fr_{\rm rms}=u_{\rm rms}/N\lambda_0$ based on the flow 
\textit{root-mean-square} velocity $u_{\rm rms}$ and on the typical wavelength 
$\lambda_0=15$~cm of the flow mode at the forcing frequency $\sigma_0$ 
(extracted from temporal Fourier filtering of the measured velocity field at 
$\sigma_0$). The \textit{root-mean-square} (rms) velocity is computed as 
$u_{\rm rms}=\langle\sqrt{\langle u_x^2 +u_z^2 \rangle_t}\rangle_{\bf x}$, 
where $\langle \, \rangle_t$ stands for the temporal average and $\langle \, 
\rangle_{\bf x}$ the spatial average over the PIV region. As we will see in the 
following section, the flow mode at the forcing frequency has a wavelength 
spectrum typically ranging from $10$ to $20$~cm. The value $\lambda_0=15$~cm 
used in Tab.~\ref{tab:freqb} to compute $Re_{\rm rms}$ and $Fr_{\rm rms}$ is 
therefore no more than an order-of-magnitude aiming at being representative of 
the forced mode. It is worth to note that the range of lengthscales of 
the forced mode observed in our experiments is consistent in order of magnitude 
with estimates from the theory of self-similar internal wave 
beams~\cite{Thomas1972, 
Machicoane2015} which are commonly produced by oscillating 
cylinders~\cite{Mowbray1967, Thomas1972, Sutherland2002, Gostiaux2007, 
Ermanyuk2008, Machicoane2015}.

For the sake of completeness, we also report in Tab.~\ref{tab:freqb} the values 
of the buoyancy Reynolds number $Re_b=Re_{\rm rms} Fr_{\rm rms}^2$. As 
mentioned in the introduction, $Re_b$ is a key parameter for the ``strongly 
stratified turbulence'' regime~\cite{Brethouwer2007,Rorai2015,Maffioli2017}. 
For this strongly non-linear regime, when $Re_b$ is large, a turbulent cascade 
from large to small scales indeed develops, whereas, when $Re_b$ is small, no 
cascading process is retrieved, and energy dissipation mainly occurs at the 
injection scale. We highlight however that the importance of the buoyancy 
Reynolds number $Re_b$ for the weakly non-linear regimes of stratified 
turbulence, for which non-linear interactions between internal waves dominate 
the flow dynamics, is yet to be established.

\section{Experimental results}

\subsection{Non-linear emergence of standing subharmonic modes}\label{sc:standingmodes}

\begin{figure}
	\centerline{\includegraphics[width=12cm]{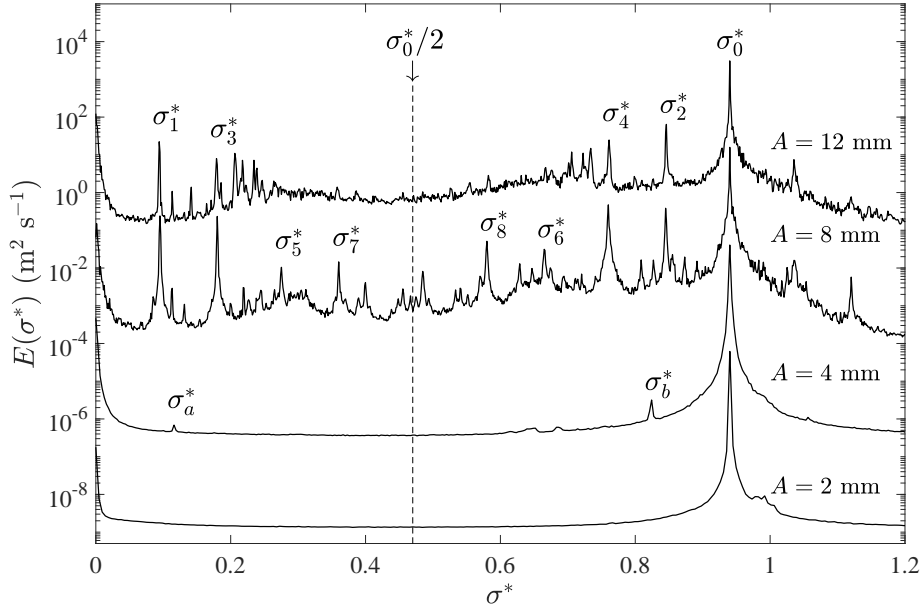}}
	\caption{Temporal power spectral density $E(\sigma^*)$ of the measured velocity field as a function of the normalized frequency $\sigma^*=\sigma/N$ for the experiments of Tab~\ref{tab:freqb}. A vertical shift by a factor of $100$ is introduced between successive spectra for better visualisation.}\label{fig_sans_bo}
\end{figure}

In this section, we report on experiments conducted at four forcing amplitudes $A$ (see Tab~\ref{tab:freqb}) aiming to explore the emergence of non-linear effects in the flow. First, in Fig.~\ref{fig_sans_bo}, we show the temporal power spectral density $E(\sigma^*)$ of the velocity field as a function of the normalized frequency $\sigma^*=\sigma/N$. Depending on the experimental run, these spectra are computed over $200$ to $400$~forcing periods $T=2\pi/\sigma_0$, in the statistically steady state of the flow. At the smallest forcing amplitude $A=2$~mm, we observe a dominant peak at the driving frequency $\sigma_0^*=0.94$, which carries almost all of the flow kinetic energy. We also observe a secondary peak at $\sigma^* = 0$ (i.e., $\sigma^*\leq 0.03$) which represents only a tiny fraction (of about $0.1\%$) of the total kinetic energy. Previous research on stratified fluids has shown~\cite{Sutherland2006,Ermanyuk2008,King2009,Bordes2012b,Scolan2013,Bourget2013,Brouzet2016,Brouzet2017,Shmakova2019,Fan2020,Davis2020,Jamin2021} that such (quasi-)steady flow could stem from streaming and Stokes drift non-linear processes, which may develop in the bulk of the flow or in the vicinity of the wavemakers. The rest of the kinetic energy (of the order of $0.1\%$) is associated to harmonics of the forced mode which is also a common feature of internal gravity wave experiments~\cite{Tabaei2005,King2009,Shmakova2017,Shmakova2019,Fan2020,Shmakova2021,Davis2020,Rodda2022}. The harmonics cannot here propagate as internal waves because their non-dimensional frequency $\sigma/N$ is larger than 1.

\begin{figure}
	\centerline{\includegraphics[width=8.5cm]{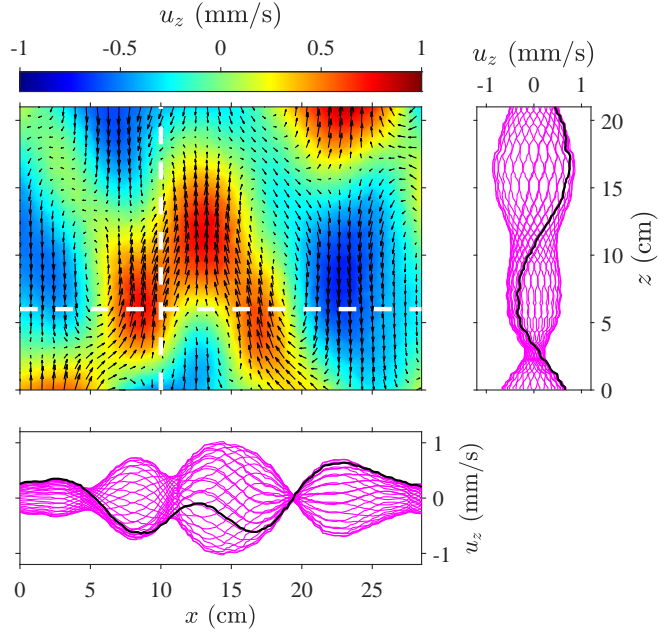}}
	\caption{Velocity field Fourier filtered at the forcing frequency $\sigma_0^*=0.94$ for the experiment at amplitude $A=2$~mm. The main figure shows a typical snapshot of the velocity field. This field is accompanied by two panels showing the temporal evolution of the profiles of the vertical velocity $u_z$ along a horizontal line (bottom panel) and along a vertical line (right panel) which are highlighted in the velocity field by dashed lines. In this figure and the following, the origin of the vertical coordinate $z$ has been redefined to match the bottom of the PIV measurement region.}\label{fig:champ4mm}
\end{figure}

In Fig.~\ref{fig:champ4mm}, we show a typical snapshot of the velocity field at 
$A=2$~mm, which we have Fourier filtered at the forcing frequency 
$\sigma_0^*=0.94$. This velocity field is accompanied by two panels showing the 
temporal evolution of the profiles of the vertical velocity $u_z$ along a 
horizontal line and a vertical line. As expected, the velocity profiles show 
that the forced flow is composed of an ensemble of traveling and coherent wave 
beams. We also note that the flow exhibits vertically elongated structures and 
that the magnitude of the vertical velocity is on average larger than that of 
the horizontal velocity. These two features are in qualitative agreement with 
the dispersion relation of internal gravity waves at the considered 
non-dimensional frequency $\sigma_0^*=0.94$ (see Ref.~\cite{Savaro2020}). To be 
more specific, the velocity snapshots of the forced mode, such as the one of 
Fig.~\ref{fig:champ4mm}, reveal typical horizontal scales $\lambda_x=2\pi/k_x$  
of the order of $15\pm 5$~cm and typical vertical scales $\lambda_z=2\pi/k_z$ 
of the order of $25\pm 5$~cm. This anisotropy in scales is consistent with the 
internal wave dispersion relation which predicts that $\sqrt{k_x^2 + 
k_y^2}/|k_z|=\sigma^*_0/(1-\sigma^{*2}_0)^{1/2} \simeq 2.8$ for waves at 
non-dimensional frequency $\sigma_0^*=0.94$. The difference between the typical 
experimental ratio $\lambda_z/\lambda_x \simeq 1.7$ and the previous 
theoretical prediction can be explained by the fact the wave beams emitted by 
the oscillating cylinders are propagating in vertical planes that make an angle 
(either of $\varphi=30^\circ$ or of $\varphi=60^\circ$ depending on the 
considered cylinder, see Fig.~\ref{fig_expset}) with the velocity measurement 
vertical plane. Thus, we expect to observe, in the velocity measurement plane, 
horizontal wavelengths larger (by a factor $1/\cos\varphi$) than that in the 
vertical planes in which the waves propagate.

Looking back to Fig.~\ref{fig_sans_bo} and increasing the forcing amplitude to $A=4$~mm, we observe the emergence of two tiny subharmonic peaks in the temporal kinetic energy spectrum, at two non-dimensional frequencies $\sigma_a^*=0.116$ and $\sigma_b^*=0.824$ which are in temporal triadic resonance with the forced mode frequency. These subharmonic modes are very weak, having an energy density four orders of magnitude smaller than that of the forced mode at $\sigma_0^*=0.94$. This prevents us from conducting a proper study of the spatial structure of the velocity fields Fourier-filtered at $\sigma_a^*$ and $\sigma_b^*$, which turns out to be polluted by measurement noise. Nevertheless, it is likely that the mechanism explaining the emergence of these two subharmonic energy peaks is the triadic resonance instability (TRI) of the forced waves at $\sigma_0$, a process that has been extensively studied theoretically, experimentally and numerically~\cite{McEwan1972, Koudella2006, Korobov2008, Joubaud2012, Bourget2013, Scolan2013, Bourget2014, Brouzet2017, Karimi2014}: The TRI drains the energy of the primary wave at frequency $\sigma_0$ and wavevector $\bf{k_0}$ toward couples of subharmonic waves at frequencies $\sigma_i$ and $\sigma_j$ and wavevectors $\bf{k_i}$ and $\bf{k_j}$ in temporal, $\sigma_i+\sigma_j = \sigma_0$, and spatial, $\pm\bf{k_i}\pm\bf{k_j}=\bf{k_0}$, resonances with the primary wave. The detailed features of the secondary waves produced by the TRI have been shown to depend on the topology of the primary wave and of the fluid domain (plane wave, finite width wave beam, standing mode, wave attractor). Previous research has also reported that the instability depends on the primary wave Reynolds number $Re$ and, in particular, that it is triggered only beyond a threshold in $Re$ in the general case. Here, in the experiment at forcing amplitude $A=4$~mm, the flow is most likely close to the onset of the triadic resonance instability.

\begin{figure}
	\centerline{\includegraphics[width=12cm]{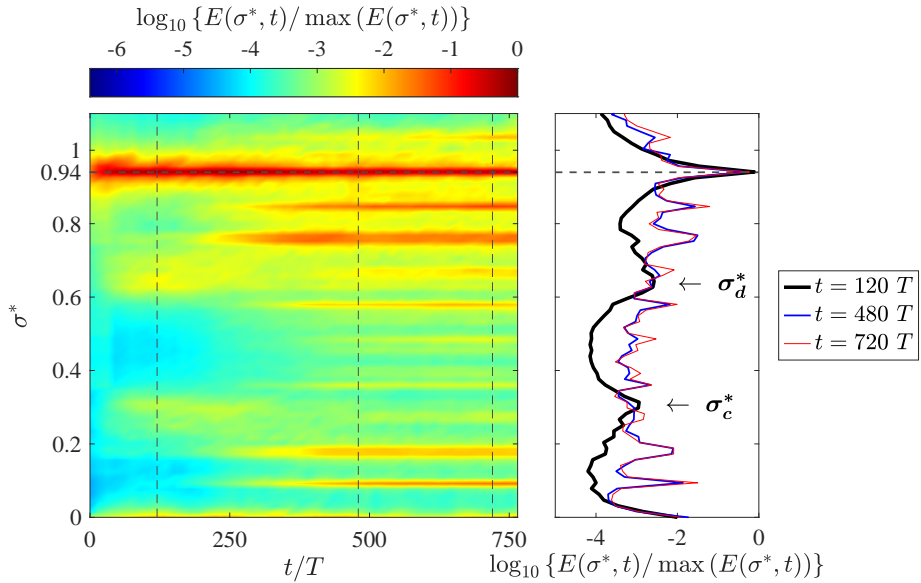}}
\caption{Time-frequency kinetic energy spectrum for the experiment at $A=8~$mm (see Tab~\ref{tab:freqb}). (left) Logarithm of the temporal energy spectra $E(\sigma,t)$ normalized by its maximum as a function of time $t$ and of the normalized frequency $\sigma^*=\sigma/N$. $t=0$ corresponds to the start of the forcing. The spectrum is computed using a sliding time window of $60\,T$. (right) Corresponding temporal energy spectra at different times, $t=120\,T,~480\,T$ and $720\,T$, highlighted by vertical dashed lines in the left panel.}\label{fig_tempfreq}
\end{figure}

We proceed by considering the experiment at forcing amplitude $A=8~$mm. Before discussing its temporal spectrum in the statistically steady state reported in Fig.~\ref{fig_sans_bo}, it is worth to study its evolution during the transient regime. To this end, we plot in the left panel of Fig.~\ref{fig_tempfreq} the chart of the temporal energy spectrum $E(\sigma,t)$ of the velocity field as a function of time $t$ and of the normalized frequency $\sigma^*=\sigma/N$. This time-frequency spectrum is computed using a sliding time window of $\Delta T = 60\,T$ where $T=2\pi/\sigma_0$ is the forcing period. In the right panel of Fig.~\ref{fig_tempfreq}, we show three cuts of this chart, corresponding to the temporal spectra at $t=120\,T,~480\,T$ and $720\,T$ after the start of the forcing. It is worth to note that the temporal spectra have here a significantly lower frequency resolution than those of Fig.~\ref{fig_sans_bo} because they are computed over short time windows of $60\,T$.

In Fig.~\ref{fig_tempfreq}, we first observe the growth of the energy peak associated to the forced mode at $\sigma_0^*=0.94$ which reaches a quasi-stationary state less than thirty forcing periods $T$ after the start of the forcing (at $t=0$). Second, shortly after the start of the forcing, we can see the growth of two wide subharmonic bumps temporally resonant with the forcing frequency and with their maxima found at subharmonic frequencies $\sigma_c^* \simeq 0.31$ and $\sigma_d^* \simeq 0.63$ (see the temporal spectrum at $t=120\,T$ in the right panel of Fig.~\ref{fig_tempfreq}). These bumps seem to reach a somewhat quasi-stationary maximum at time $t\simeq 80\, T$ before they start to decrease around $t = 250\,T$.

The emergence of these two wide subharmonic energy bumps is reminiscent of the scenario of the triadic resonance instability observed for inertial waves in rotating fluids~\cite{Bordes2012,Brunet2019,Monsalve2020,Mora2021}. In the case of rotating fluid experiments, this ``two-wide-subharmonic-bumps'' state corresponds to the steady state of the flow when the forcing Reynolds number is moderately larger than the threshold of the TRI. Moreover, it has been shown that, when the Reynolds number is further increased, this state can progressively transform into a continuum of energy in the inertial wave frequency domain whose features are in quantitative agreement with the Weak Turbulence Theory for inertial waves in rotating fluids~\cite{Monsalve2020}.

Here, contrary to the case of rotating fluids, the two wide subharmonic bumps have disappeared in the spectrum in the steady state. In Fig.~\ref{fig_tempfreq}, we indeed see that $300\,T$ after the start of the forcing, the wide subharmonic bumps progressively vanish while an ensemble of sharp subharmonic peaks progressively establishes over typically the following $200\,T$. This regime characterized by several sharp peaks in the temporal energy spectrum seems to be the steady state of the flow and corresponds to the spectrum reported in Fig.~\ref{fig_sans_bo}. The physical process behind this transition of the non-linear state of the flow is an open question. However, as discussed in the following, it reveals the attraction of the subharmonic energy to discrete resonance frequencies associated to eigenmodes of the fluid domain.

\begin{table}
\caption{Normalized frequencies $\sigma_i^*$ associated to the dominant peaks found in the temporal energy spectrum of the experiment at $A=8$~mm in Fig.~\ref{fig_sans_bo}.}\label{tab:freqbsig}
\begin{ruledtabular}
\begin{tabular}{ c c c c c c c c c}
\hfill
$\sigma^*_1$ & $\sigma^*_2$ & $\sigma^*_3$ & $\sigma^*_4$ & $\sigma^*_5$& $\sigma^*_6$& $\sigma^*_7$ & $\sigma^*_8$ & $\sigma^*_0$\\
\hline
0.094 & 0.846 & 0.179 &0.761& 0.276  & 0.665 & 0.361 & 0.580& 0.940\\
\end{tabular}
\end{ruledtabular}
\end{table}

Returning to the analysis of the statistically steady state of the flow (see 
Fig.~\ref{fig_sans_bo}), the experiment at $A=8~$mm still exhibits a dominant 
peak at the forcing frequency $\sigma_0^*=0.94$ in its temporal kinetic energy 
spectrum. However, as already said, several other peaks emerged at specific 
subharmonic frequencies ($\sigma^*<\sigma_0^*$), which are labeled as 
$\sigma_i^*$ with $i=1, 2, 3 \dots$ (see Fig.~\ref{fig_sans_bo}). Besides these 
sharp peaks, the spectrum at $A=8$~mm also reveals the presence, over the whole 
subharmonic frequency range, of a weak but continuous energy background, whose 
energy content slowly increases as the frequency $\sigma^*$ increases. Finally, 
for frequencies beyond the peak at $\sigma^*_0=0.94$, the energy spectrum 
background rapidly decays in amplitude, with the noticeable presence of weakly 
energetic peaks.

A closer inspection of the temporal energy spectrum in the subharmonic range ($\sigma^*<\sigma_0^*$) shows that each sharp peak with a frequency below $\sigma_0^*/2$ has a twin peak at the symmetric frequency with respect to $\sigma_0^*/2$ (see Tab.~\ref{tab:freqbsig}). This result advances that these couples of peaks are in temporal triadic resonance with the forced mode. 
For instance, it can easily be verified in Fig.~\ref{fig_sans_bo} (and also in Tab.~\ref{tab:freqbsig}) that $\sigma_1+\sigma_2=\sigma_0$, $\sigma_3+\sigma_4=\sigma_0$, $\sigma_5+\sigma_6=\sigma_0$, $\sigma_7+\sigma_8=\sigma_0$.
The temporal spectrum, however, reveals an even richer dynamics, as we observe that these subharmonic peaks fulfill additional temporal triadic resonance conditions within themselves, e.g., $\sigma_1+\sigma_3=\sigma_5$, $\sigma_1+\sigma_6 = \sigma_4$, $\sigma_3+\sigma_6 = \sigma_2$, $\sigma_3 + \sigma_3 = \sigma_7$, $\sigma_3 + \sigma_8 = \sigma_4$. On the one hand, it is likely that the primary mechanism at the origin of (at least some of) the subharmonic peaks is the triadic resonance instability of the waves at the forcing frequency $\sigma_0$~\cite{McEwan1972, Koudella2006, Joubaud2012, Bourget2013, Scolan2013, Bourget2014, Brouzet2017, Karimi2014}, in a regime where it couples to the flow boundaries to select specific discrete resonance frequencies. On the other hand, and as already reported in previous experimental works~\cite{Brouzet2016,Brouzet2017,Davis2019,Savaro2020,Rodda2022}, the observed richness in temporal triadic resonances necessarily implies that additional triadic interactions are at play. This could be the TRI of secondary waves, the non-linear interactions of two modes yielding a third mode~\cite{Korobov2008,Dobra2022} or the growth of harmonics of a mode (e.g. the mode at $\sigma_7=2\sigma_3$)~\cite{Thorpe1968,Korobov2008,Rodda2022}. Finally, it is also worth to mention that recent experiments~\cite{Davis2019,Rodda2022} have shown that triadic interactions between two discrete wave modes may also lead, under certain conditions, to the forcing of oscillating modes which do not follow the dispersion relation. In Ref.~\cite{Rodda2022}, these modes are called \textit{bound waves} by analogy with processes observed for surface waves~\cite{Herbert2010}.

The steady state observed in our experiments with a temporal spectrum dominated by subharmonic discrete energy peaks is reminiscent of previous experiments in which internal waves are forced at a specific frequency and at large scale~\cite{Brouzet2016,Brouzet2017,Davis2019,Savaro2020,Davis2020}. Interestingly, in two of these works~\cite{Brouzet2017,Savaro2020}, the authors report that some of the spectral energy peaks correspond to standing wave modes of the fluid domain. For a parallelepipedic fluid domain with a square base, which is the geometry relevant to our experiments, standing eigenmodes of internal gravity waves have a simple spatio-temporal structure~\cite{Maas2003,Savaro2020} 
\begin{eqnarray}
    u_x& \propto & \sin(k_x x) \cos(k_y y) \cos(k_z z) \cos(\sigma t)\, , \label{eq:modex}\\
    u_y& \propto & \cos(k_x x) \sin(k_y y) \cos(k_z z) \cos(\sigma t)\, , \label{eq:modey}\\
    u_z& \propto & \cos(k_x x) \cos(k_y y) \sin(k_z z) \cos(\sigma t)\, , \label{eq:modez}
\end{eqnarray}
with the wavevector components taking discrete values $k_x=\pi n_x/L$, $k_y=\pi n_y/L$ and $k_z=\pi n_z/H$. Here, $L$ is the side length of the square base and $H$ the height of the fluid domain, and, $n_x$, $n_y$ and $n_z$ take integer values accounting for the number of half-wavelengths in each cartesian direction. The standing modes are found for discrete frequencies verifying the internal wave dispersion relation~~\cite{Maas2003,Savaro2020}
\begin{equation}\label{eq:disp_discrete}
    \sigma^{*2}=\frac{(n_x^2+n_y^2)/L^2}{(n_x^2+n_y^2)/L^2+n_z^2/H^2}\, .
\end{equation}

\begin{figure}
		\centerline{\includegraphics[width=14cm]{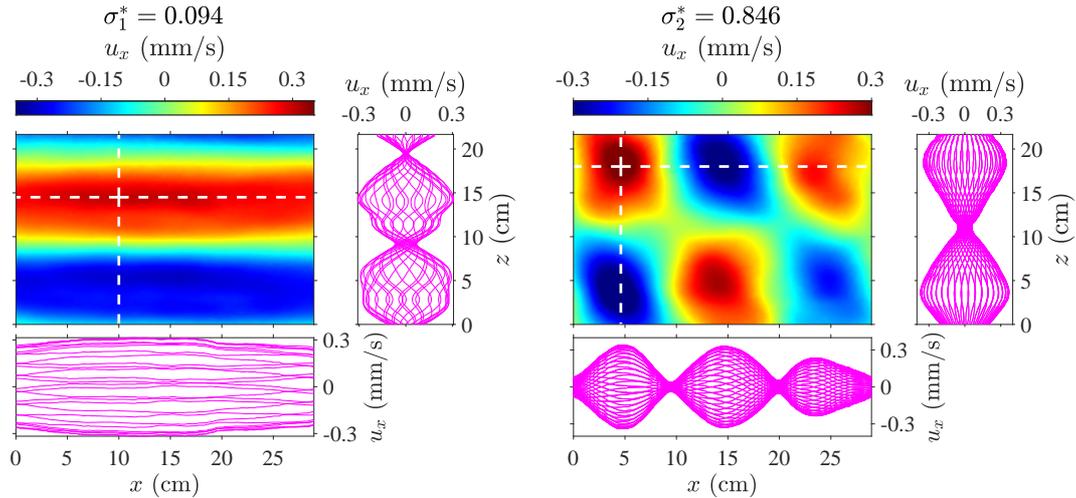}}
	\caption{(left) Typical snapshot of the horizontal velocity $u_x$ Fourier filtered at frequency $\sigma_1^*= 0.094$ for the experiment at amplitude $A=8$~mm. The field is accompanied by two panels showing the temporal evolution of the profile of the horizontal velocity $u_x$ along a horizontal line (bottom panel) and along a vertical line (right panel) which are highlighted in the main figure by dashed lines. (right) Same panels for the velocity field Fourier filtered at $\sigma_2^*= 0.846$.}\label{fig:champsig}
\end{figure}

In order to assess the presence of standing modes in our experiments, we report in Fig.~\ref{fig:champsig} typical snapshots of the horizontal velocity field at $A=8~$mm filtered at the peak frequencies $\sigma_1^*= 0.094$ (left) and $\sigma_2^*= 0.846$ (right), which are in temporal triadic resonance with the forced waves. As in Fig.~\ref{fig:champ4mm}, we plot for each frequency, in addition to the velocity field, the temporal evolution of the velocity profiles along a specific horizontal line and a specific vertical line. From the left panels, we observe that the velocity field at $\sigma_1^*= 0.094$ exhibits a large scale horizontally elongated structure with a typical vertical wavelength of the order of $20$~cm. At first sight, the presence of nodes and antinodes in the velocity profiles of Fig.~\ref{fig:champsig}(left) suggests that the peak at $\sigma_1^*$ is dominated by a standing wave, which correlates well with the structure of a fluid domain eigenmode of indices $n_x=1$ and $n_z=6$. Assuming $n_y=0$ and injecting these index values into the theoretical frequency expression (\ref{eq:disp_discrete}) we recover $\sigma^* \simeq 0.098$, which is in good agreement with the experimental peak frequency $\sigma_1^*= 0.094$. Although the index value $n_y=0$ cannot be directly confirmed from our PIV measurements, which do not give us access to the flow structure in the $y$ direction, picking other values for $n_y$ leads to frequencies in clear discrepancy with the experimental peak frequency $\sigma^*_1$: for instance, $n_y=1$ leads to $\sigma^*\simeq 0.138$ and $n_y=2$ to $\sigma^*\simeq 0.215$. Thus, it seems reasonable to consider that the peak at $\sigma_1^*= 0.094$ is dominated by the fluid domain eigenmode ($n_x^{(1)}=1, n_y^{(1)}=0, n_z^{(1)}=6$). It is worth to note that this agreement between our data and the eigenmode theory for a parallelepiped is rather satisfactory considering that the presence of the oscillating cylinders and the bars holding them renders our fluid domain more complex than a parallelepiped.

Likewise, we find evidence suggesting that the peak at $\sigma_2^*= 0.846$ is associated to an eigenmode of the fluid domain. The right panels of Fig.~\ref{fig:champsig} indeed reveal a checkerboard pattern with regular nodes and antinodes. This mode has a spatial structure compatible with indices $n_x^{(2)}=10$ and $n_z^{(2)}=4$, but contrary to the mode at $\sigma_1^*$, it is unclear which index to pick for $n_y$: for instance, injecting $n_y^{(2)}=3, 4$ or $5$ (along with $n_x^{(2)}=10$ and $n_z^{(2)}=4$) into Eq.~(\ref{eq:disp_discrete}) yields $\sigma^* \simeq 0.839,0.846$ or $0.855$ respectively, values which are all close to the experimental peak frequency $\sigma_2^*$. 

\begin{figure}
	\centerline{\includegraphics[width=14cm]{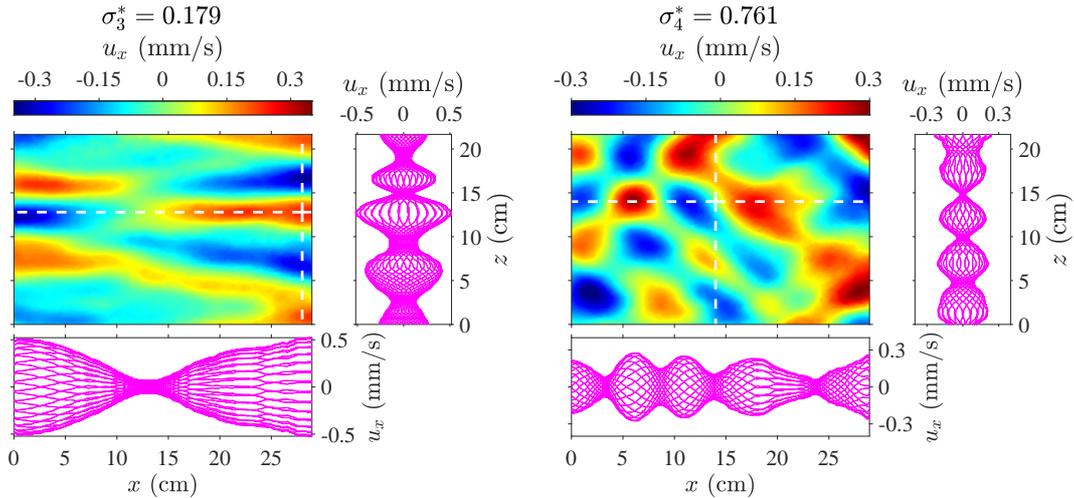}}
 	\caption{(left) Typical snapshot of the horizontal velocity $u_x$ Fourier filtered at frequency $\sigma_3^*= 0.179$ for the experiment at amplitude $A=8$~mm. The field is accompanied by two panels showing the temporal evolution of the profile of the horizontal velocity $u_x$ along a horizontal line (bottom panel) and along a vertical line (right panel) which are highlighted in the main figure by dashed lines. (right) Same panels for the velocity field Fourier filtered at $\sigma_4^*=0.761$.}\label{fig:champsig34}
\end{figure}

The analysis we presented for the couple of subharmonic peaks at $(\sigma_1^*,\sigma_2^*)$ can be conducted for the other couples of peaks highlighted in Fig.~\ref{fig_sans_bo}, which are in temporal resonance with the forced waves. For the couple of peaks at frequencies $\sigma_3^*= 0.179$ and $\sigma_4^*= 0.761$, the velocity fields and profiles shown in Fig.~\ref{fig:champsig34} reveal structures roughly compatible with eigenmodes of the fluid domain. In fact, as we consider less energetic peaks in the series of $\sigma_i^*$ (with $i$ from $3$ to $8$), we observe that the node-antinode structure of the velocity fields becomes less regular, a feature that can already be seen for the couple $(\sigma_3^*,\sigma_4^*)$. This observation might reveal perturbations of the standing eigenmodes by waves associated to the continuous energy background, and whose influence grows relatively as we consider less energetic peaks. Nevertheless, as stated above, it is also possible that some of the subharmonic energy peaks in the spectrum at $A=8$~mm might not directly be associated to eigenmodes but result from the non-linear interaction between two discrete modes.

Next, we consider the experiment at forcing amplitude $A=12$~mm. Although not shown here, the transient regime of the flow is quite similar to that of the experiment at $A=8$~mm (illustrated in Fig.~\ref{fig_tempfreq}): First, two wide subharmonic spectral bumps emerge before vanishing around $t=300\,T$ after the start of the forcing, while several sharp energy peaks grow until saturation to build the steady state of the flow. In the temporal spectrum in the steady regime (Fig.~\ref{fig_sans_bo}), we first observe that the dominant peak couples at the frequencies $\sigma_i^*$ with $i=1,2,3,4$ seen at $A=8$~mm are still present at $A=12$~mm. On the contrary, the peaks associated to frequency couples $(\sigma_5^*,\sigma_6^*)$ and $(\sigma_7^*,\sigma_8^*)$ are no longer observable at $A=12$~mm, whereas new energetically subdominant peaks are found in the frequency range between $\sigma_3^*$ and $\sigma_4^*$. In parallel, the continuous background energy density observed in the frequency range $\sigma^*\leq \sigma_0^*$ becomes more energetic at $A=12$~mm than at $A=8$~mm relatively to the discrete sharp peaks. 
These observations are comparable with those of Savaro \textit{et 
al.}~\cite{Savaro2020} who propose that the background continuum of energy in 
frequency results from a random ensemble of traveling internal waves.

\subsection{Discussion}\label{sec:discussion}

In summary, the results of the previous section mainly show that the non-linear statistically steady flow produced by our setup is composed, in addition to the internal waves at the forcing frequency $\sigma_0$, of multiple wave modes found at discrete subharmonic frequencies (i.e., lower than $\sigma_0$). These modes are involved in several triadic resonances with the forced mode and between themselves. Our analysis suggests that they are wave eigenmodes of the fluid domain, at least for the most energetic of them. This scenario is consistent with several previous experimental studies aiming to produce an internal wave turbulence by forcing waves at a specific frequency~\cite{Brouzet2016,Brouzet2017,Davis2019,Savaro2020,Davis2020}.

Besides, we find that the transient regime between the start of the forcing and the establishment of the flow steady state is more complex than a simple progressive growth of the discrete subharmonic energy peaks in the temporal spectrum. Indeed, during a first transient stage, which typically lasts several hundreds of forcing periods, the time-frequency spectrum reveals the growth of two wide subharmonic bumps in temporal resonance with the forced waves. These bumps later vanish while the eigenmodes associated to the steady state energy peaks slowly establish. A similar scenario has been reported during the transient of non-linear internal gravity wave attractor experiments in Refs.~\cite{Brouzet2016,Brouzet2017,Davis2019,Davis2020}. In Ref.~\cite{Brouzet2017}, the initial transient subharmonic modes are moreover shown to be spatially localized internal waves produced by the TRI of the forced waves. 

Nevertheless, our observation at early times of two relatively wide subharmonic bumps is somewhat different from previous experimental studies, which have overall reported, at early times, couples of transient subharmonic modes associated to sharp energy peaks~\cite{Bourget2013,Scolan2013,Brouzet2016,Brouzet2017,Davis2019,Davis2020,Fan2020}. This difference with our observations might be related to the fact our forced mode is nearly homogeneous and statistically axisymetric whereas the cited works involve quasi-2D (except in \cite{Fan2020}) and localized forced waves.

In a way similar to Savaro~\textit{et al.}~\cite{Savaro2020}, the standing 
eigenmodes coexist in our experiments with a weakly energetic background which 
is continuous in frequency and found in the subharmonic range. According to 
Savaro~\textit{et al.}~\cite{Savaro2020}, this energy background is associated 
to a random ensemble of traveling internal gravity waves. In our experiments as 
well as in Ref.~\cite{Savaro2020}, the level of this continuum of energy in 
frequency grows more rapidly than that of the eigenmodes when the forcing 
amplitude is increased. This suggests that, as we increase the Reynolds number, 
the flow is starting to transition toward a wave turbulence-like regime, 
characterized by a continuum of energy in the internal wave frequency domain 
and by weak finite-size effects. 
It is worth to note that the competition between traveling modes and 
standing modes of internal gravity waves has been proposed to be relevant in 
astrophysical systems such as in the radiative zone of 
stars~\cite{Alvan2014} and in 
stratified planetary cores~\cite{Bouffard2022}.

As stated in the introduction, our original purpose is to achieve a weakly-nonlinear internal wave turbulence regime approaching the assumptions under which the weak turbulence theory (WTT) applied to stratified fluids has been derived~\cite{Caillol2000,Lvov2004,Lvov2010,Nazarenko2011,Dematteis2021}. A key assumption is the infinite domain limit, which implies that the scales of the waves involved in the energy cascade are small compared to the system size, and that no domain eigenmodes leading to a discretization of the energy in the frequency and wavenumber spaces emerge.

Given that in our experiments, as the forcing Reynolds number is increased, we observe that the spectral peaks associated to the eigenmodes are progressively engulfed by an increasing energy continuum in frequency, a natural way forward is to explore even larger forcing amplitudes. However, as discussed in Sec.~\ref{sec:setup}, further increasing the forcing amplitude results in the onset of significant irreversible mixing of the fluid stratification hinting the emergence of strong non-linearities. Since we want to restrict our study to weakly non-linear regimes in a fluid with a steady linear and stable stratification, we abstain from further increasing the forcing amplitude and leave the study of the mixing processes to future works.

An alternative strategy to approach a wave turbulence regime would be to inhibit the emergence of the fluid domain eigenmodes so that the transient initial non-linear regime that we evidenced, hopefully composed of random propagative waves, becomes the steady state of the flow. An attempt to achieve this goal is presented in the next section.

\subsection{One step closer to the internal wave turbulence regime}\label{sec:tiltedplanes}

\begin{figure}
 	\centerline{\includegraphics[width=13cm]{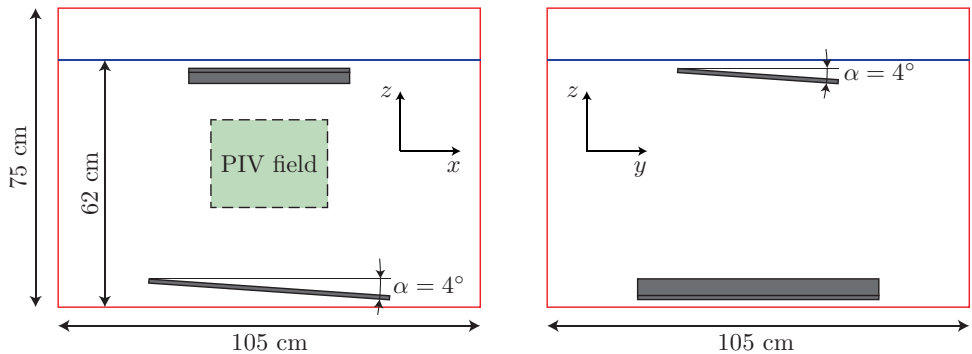}}
    \caption{Sketch of the modification of the experimental setup. We introduced two square panels inclined at an angle $\alpha=4^\circ$ with respect to the horizontal. The two panels are horizontally centered in the fluid domain, one at the top and one at the bottom. The top panel, of $40 \times 40$~cm$^2$, and the bottom panel, of $60 \times 60$~cm$^2$, are both placed in the region where the fluid is stratified, i.e. outside of the $2$ to $3$~cm mixed layers at the very top and bottom of the fluid domain. The tilt directions of the two panels in the horizontal plane are normal to each other: the bottom panel is tilted in the $x$-direction whereas the top panel is tilted in the $y$-direction. The oscillating cylinders are not shown here for sake of clarity. Their spatial arrangement has not been modified and is identical to that of Fig.~\ref{fig_expset}.}\label{fig_expset2}
\end{figure}

In the following, we report experiments (see Tab.~\ref{tab:freqb2}) conducted using the same configuration as for the experiments presented in Sec.~\ref{sc:standingmodes}. We however introduced one important modification: at the top and at the bottom of the fluid domain, we added rectangular panels inclined at an angle $\alpha=4^\circ$ with respect to the horizontal. This new experimental configuration is sketched in Fig.~\ref{fig_expset2}. The two panels are horizontally centered in the fluid domain and the directions in the horizontal plane of their respective tilt are normal to each other. The top panel surface is of $40 \times 40$~cm$^2$ and the one of the bottom panel of $60 \times 60$~cm$^2$.

\begin{table}
\caption{Parameters of the experiments ``with tilted planes'' reported in Sec.~\ref{sec:tiltedplanes}. $A$ is the cylinders amplitude of motion. $Re_f=A\sigma_0 d/\nu$ and $Fr_f=A\sigma_0/Nd$ are the forcing Reynolds and Froude numbers, $Re_{\rm rms}=u_{\rm rms}\lambda_0/\nu$, $Fr_{\rm rms}=u_{\rm rms}/N\lambda_0$ and $Re_b=Re_{\rm rms} Fr_{\rm rms}^2$, the Reynolds, Froude and buoyancy Reynolds numbers based on the flow rms velocity $u_{\rm rms}$ and $\lambda_0=15$~cm, the typical wavelength of the flow mode at the forcing frequency $\sigma_0=0.94\,N$. $d=4$~cm is the cylinders diameter, $\nu=10^{-6}$~m$^2$/s the average kinematic viscosity of the fluid and $N=1$~rad/s the average buoyancy frequency.} 
\label{tab:freqb2}
\begin{ruledtabular}
\begin{tabular}{c c c c c c c}
$A$ (mm)  &  $Re_f$ & $Fr_f$  & $u_{\rm rms}$ (mm/s) & $Re_{\rm rms}$ & $Fr_{\rm rms}$ & $Re_b$ \\
\hline
2 & 75 & 0.05 & 0.9 & 135 & 0.006 & 0.005 \\
4  & 150 & 0.09 & 1.7 & 255 & 0.011 & 0.033 \\
6 & 226 & 0.14 & 2.5 & 375 & 0.017 & 0.104 \\
8  & 301 & 0.19 & 3.1 & 465 & 0.021 & 0.199 \\
12  & 451 & 0.28 & 3.9 & 585 & 0.026 & 0.395 \\
16 & 602 & 0.38 & 4.5 & 675 & 0.030 & 0.608 \\
\end{tabular}
\end{ruledtabular}
\end{table}

Internal gravity waves follow anomalous reflection laws on inclined solid boundaries and their wavelength is modified during reflection (except for vertical and horizontal walls)~\cite{Phillips1966,Dauxois1999}. This is a consequence of their peculiar dispersion relation $\sigma^*=\sin \theta$, which relates the wave non-dimensional frequency $\sigma^*$ to the angle $\theta$ between the wave group velocity, along which the energy propagates, and the horizontal. For a reflection on a sloping wall, the ratio of the wavelengths of an incident wave and its reflected wave is equal either to $\gamma$ or $1/\gamma$ (depending on the fact the wave is going down or up the slope) with 
\begin{equation}\label{eq:gamma}
   \gamma=\left|\frac{\sin(\theta-\alpha)}{\sin(\theta+\alpha)}\right|,
\end{equation}
where $\alpha$ is the angle of the sloping wall with the horizontal.

By introducing two tilted planes with crossed orientations, we aim to 
distinguish the geometry of our fluid domain enough from a box with only 
vertical and horizontal walls to prevent the emergence of standing 
eigenmodes~\cite{Maas1997,Maas2003}. Nevertheless, at the same time, we select 
a relatively weak tilt angle, $\alpha = 4 ^\circ$, in order to limit the 
contribution of the tilted planes to the energy transfers between spatial 
scales during reflections (driven in this case by a linear process). From 
Eq.~(\ref{eq:gamma}), we can see that the wavelength modification factor 
$\gamma$ during a reflection on a tilted plane depends on the frequency of the 
wave. At the largest non-dimensional frequency in the wave domain $\sigma^*=1$, 
there is no wavelength modification and $\gamma=1$, whereas at $\sigma^*=\sin 
\alpha$ (i.e. $\theta=\alpha$) the wavelength change during a reflection is 
dramatic with $\gamma=0$. For our experimental configuration where 
$\alpha=4^\circ$, this critical reflection condition is observed for 
$\sigma^*=\sin \alpha \simeq 0.07$. We note, however, that with our choice for 
a weak angle $\alpha$, the change of wavelength during a reflection remains 
moderate over most of the wave frequency range. 

It is worth to note that fluid domains provided with a (single) tilted 
plane have also been considered in previous works, in order to trigger the 
emergence of attractors of internal or inertial 
waves~\cite{Maas1997,Grisouard2008,Scolan2013,Brouzet2017,Pillet2018,Brunet2019,Davis2020}.
At variance with our configuration, the tilt angle $\alpha$ of the plane in 
these studies is large, except in Ref.~\cite{Pillet2018} where it is 
smaller than the forced wave angle $\theta$, but still not very small 
($23^\circ$).

\begin{figure}
	\centerline{\includegraphics[width=12cm]{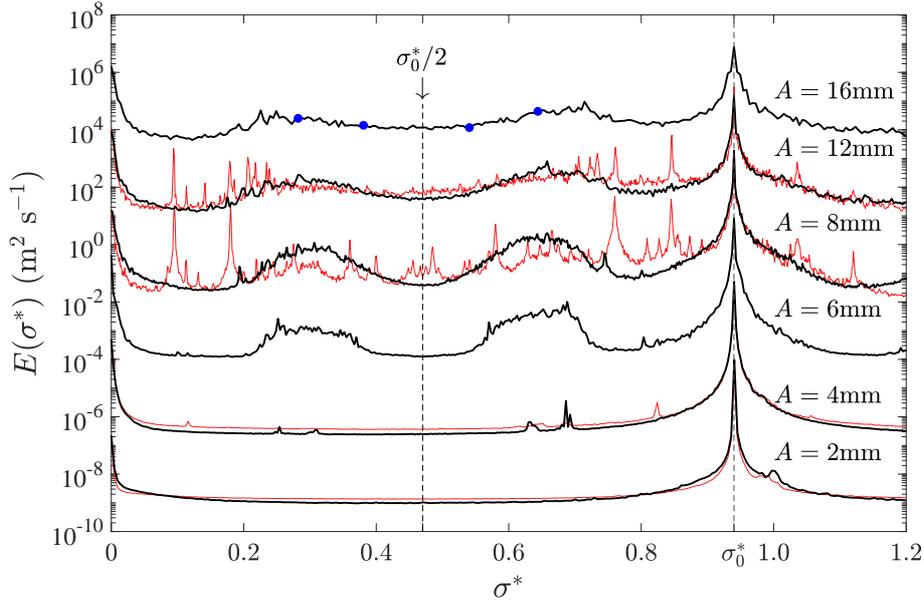}}
	\caption{Temporal power spectral density of the velocity field $E(\sigma^*)$ as a function of the normalized frequency $\sigma^*=\sigma/N$ for the experiments with the tilted planes at several forcing amplitudes $A$ (see Tab.~\ref{tab:freqb2}). A vertical shift by a factor of $100$ is introduced between successive spectra. We also report from Fig.~\ref{fig_sans_bo} the spectra of the experiments ``without tilted plane''. The spectra for the new experimental configuration ``with tilted planes'' are plotted as black thick lines whereas those of the previous series are reported as red thin lines. }\label{fig_avec_bo}
\end{figure}

In Fig.~\ref{fig_avec_bo}, we show the temporal power spectral density $E(\sigma^*)$ of the velocity field as a function of the normalized frequency $\sigma^*=\sigma/N$ for the experiments with the tilted planes (Tab.~\ref{tab:freqb2}). For comparison, we also report the spectra (of Fig.~\ref{fig_sans_bo}) for the experiments in the original configuration without tilted plane. The comparison between both datasets reveals that the spectra of the tilted plane experiments exhibit a dominant peak at the driving frequency $\sigma_0^*=0.94$ with a comparable amplitude to that of the experiments without tilted plane. Also, the energy peak associated to the quasi-steady flow at $\sigma^*\leq 0.05$ (and its evolution with the forcing amplitude) is alike for both series, with and without tilted planes. On the contrary, the energy spectra of the non-linear experiments with tilted planes at $A \ge 6$~mm show a remarkably different behavior over the subharmonic frequency range $\sigma^*<\sigma^*_0$: The discrete sharp peaks associated to eigenmodes of the fluid domain and their interactions have almost completely disappeared. They have actually been replaced by a couple of wide subharmonic bumps in temporal triadic resonance with the forced mode at $\sigma^*_0$. These subharmonic bumps are very similar to those observed during the early transient of the experiments without tilted plane. Their spectral width in non-dimensional frequency broadens from about $0.15$ to $0.30$ as the forcing amplitude increases from $A=6$~mm to $12$~mm. By further increasing the forcing amplitude to $A=16$~mm, the subharmonic bumps almost completely transform into a continuum of energy in frequency, spread over the whole subharmonic range $\sigma^*<\sigma^*_0$. 

For the sake of completeness, let us briefly discuss the behavior of our 
temporal energy spectrum at frequencies (slightly) larger than the forcing 
frequency 
$\sigma_0=0.94\,N$. We already mentioned that the recent works of Le Reun 
\textit{et 
al.}~\cite{LeReun2018} and Rodda \textit{et al.}~\cite{Rodda2022} revealed 
temporal spectra decaying as a power-law of exponent $-2$ at frequencies larger 
than the forcing frequency (a behavior extending up to $N$ in~\cite{LeReun2018}
and to typically $2N$ in~\cite{Rodda2022}). On the contrary, the decay of the 
temporal kinetic energy 
spectrum observed at $\sigma>\sigma_0$ in our experiments does not resemble a 
power-law and, if we nevertheless try to fit one over the range 
$\sigma_0<\sigma<2N$, is best fitted by a power-law decay exponent around $-4$ 
(estimates for $A=12$ and $16$~mm, with or without tilted planes).

\begin{figure}
	\centerline{\includegraphics[width=14cm]{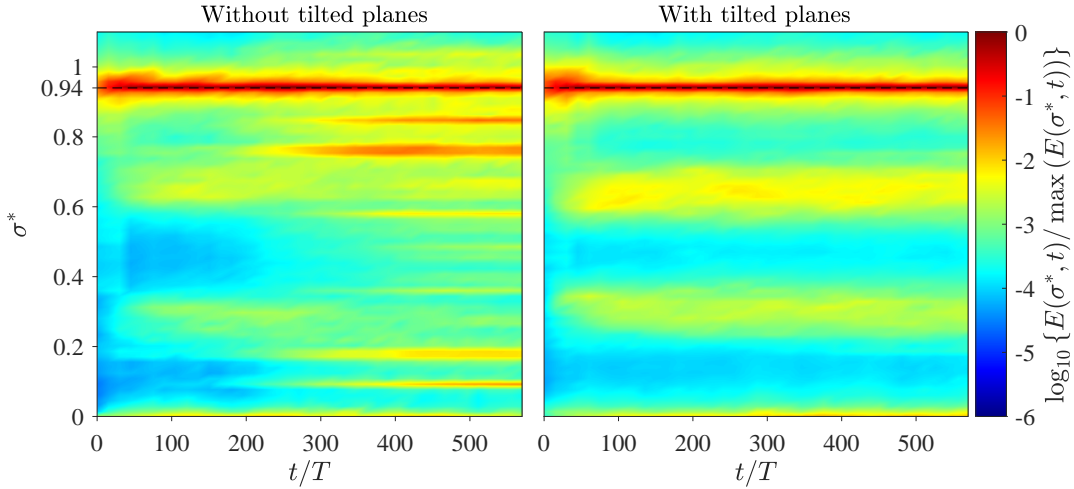}}
	\caption{Logarithm of the time-frequency kinetic energy spectrum $E(\sigma^*,t)$ for the experiments at $A=8~$mm, without tilted plane (left panel, same figure as Fig.~\ref{fig_tempfreq}) and with tilted planes (right panel). The time-frequency spectra are normalized by their maximum as a function of time $t$ and of the normalized frequency $\sigma^*=\sigma/N$. $t=0$ corresponds to the start of the forcing. The spectra are computed using a sliding time window of $60\,T$.}\label{fig:tempsFreqPI}
\end{figure}

In Fig.~\ref{fig:tempsFreqPI}, we show the time-frequency spectra of the measured velocity field for the two experiments at $A=8$~mm without tilted plane (on the left) and with tilted planes (on the right). These spectra are computed using a sliding time window of $60\,T$, where $T=2\pi/\sigma_0$ is the forcing period (the left panel is identical to Fig.~\ref{fig_tempfreq}). During the first $150\,T$ after the start of the forcing (at $t=0$), the transient regime is very similar for the two experiments, with the progressive growth of two wide bumps of energy around subharmonic frequencies, $\sigma_c^*\sim 0.3$ and $\sigma_d^*\sim 0.65$, in resonance with the forcing frequency $\sigma_0^*=0.94$. As already noticed in the discussion of Fig.~\ref{fig_tempfreq}, the wide subharmonic bumps progressively disappear beyond $t=200\,T$ for the experiment without tilted plane (left panel), while sharp peaks associated to eigenmodes slowly grow until saturation. On the contrary, for the experiment with the tilted planes (right panel), the couple of subharmonic bumps saturates in amplitude around $t=150\,T$ and then seems to correspond to the statistically steady state of the flow.

In conclusion, Figs.~\ref{fig_avec_bo} and~\ref{fig:tempsFreqPI} show that the introduction of two slightly tilted planes at the top and at the bottom of our fluid domain successfully prevents the establishment of energetic standing eigenmodes found at discrete subharmonic frequencies. As a consequence, the non-linear flow state involving two wide subharmonic bumps, which is only transient in the experiments without tilted plane, becomes the statistically stationary state in the experiments with tilted planes. This efficiency of the tilted planes to prevent the emergence of fluid domain eigenmodes is observed for all the non-linear experiments of the series with tilted planes.

\begin{figure}
	\centerline{\includegraphics[width=12cm]{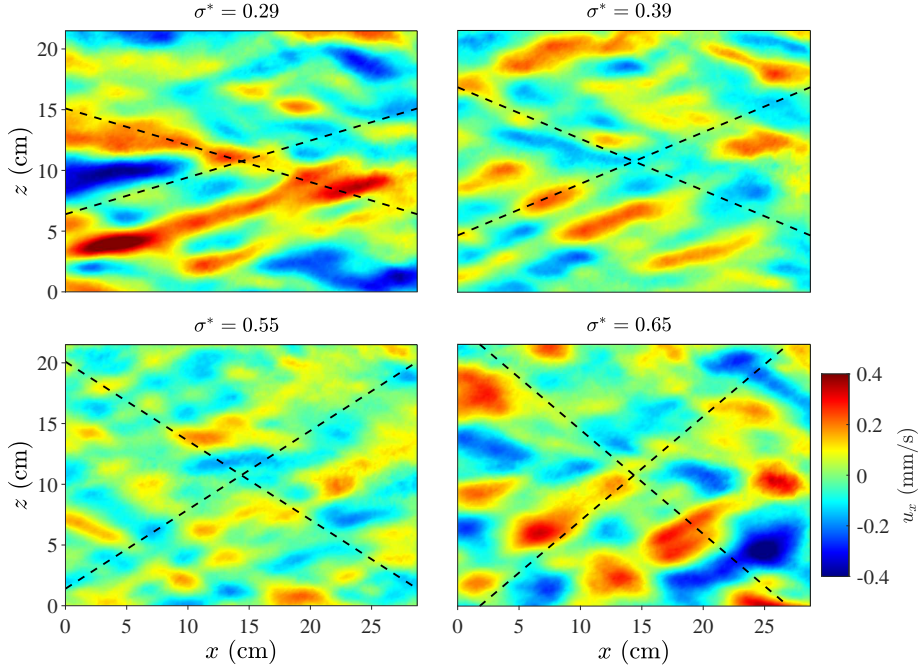}}
 	\caption{Snapshots of the horizontal component $u_x$ of the velocity field Fourier filtered at frequencies $\sigma^*= 0.29, 0.39, 0.55$ and $0.65$, for the experiment with tilted planes at amplitude $A=16$~mm. In each panel, the dashed lines are tilted of the angles $\theta=\sin^{-1}(\sigma^*)$ with respect to the horizontal. These lines indicate the directions along which internal waves at frequency $\sigma^*$, and with their wavevector in the measurement plane, propagate.}\label{fig:champsig_tilted1}
\end{figure}

We now focus on the experiment with tilted planes at the largest forcing amplitude $A=16$~mm, whose features suggest that the flow could be in a wave turbulence regime. For this experiment, we indeed observe a continuum of energy over almost the whole subharmonic frequency range $\sigma^*<\sigma^*_0=0.94$ in conjunction with the absence of discrete sharp energy peaks (Fig.~\ref{fig_avec_bo}). In the sequel, we try to assess the validity of this conjecture by characterizing the nature of the flow modes observed in the subharmonic continuum of energy of the experiment at $A=16$~mm. In Fig.~\ref{fig:champsig_tilted1}, we report snapshots of the velocity field Fourier filtered at four frequencies in the subharmonic range, $\sigma^*=0.29, 0.39, 0.55$ and $0.65$. In each of these velocity fields, a preferential tilt angle, which is increasing with the considered non-dimensional frequency, seems to dominate the flow structure. The dispersion relation of internal gravity waves, $\sigma^{* 2}=(k_x^2+k_y^2)/(k_x^2+k_y^2+k_z^2)$, actually predicts that waves at frequency $\sigma^*$ will be invariant in a direction tilted by the angle $\theta=\pm \sin^{-1}(\sigma^*)$ with respect to the horizontal (which direction is not necessarily in the velocity field measurement plane). As a guide for the eye, we therefore added two dashed lines inclined by this theoretical angle $\theta$ in each snapshot of Fig.~\ref{fig:champsig_tilted1}. These lines indicate the direction along which waves at frequency $\sigma^*$ and propagating in the measurement plane are expected to be invariant. A good agreement is found in Fig.~\ref{fig:champsig_tilted1} between the tilt shown by the theoretical dashed lines and the one of the experimental planes of constant phase.

In aggregate, the velocity fields of Fig.~\ref{fig:champsig_tilted1} are compatible with an ensemble of internal waves with a nearly spatially homogeneous statistics. Besides, we do not identify here any spatially regular structure reminiscent of standing wave modes. The wavelengths of the subharmonic modes shown in Fig.~\ref{fig:champsig_tilted1} typically range from $3$ to $10$~cm. These lengthscales are significantly smaller than the wavelengths of the forced mode and of the eigenmodes of the fluid domain observed in the experiments without tilted planes (reported in Sec.~\ref{sc:standingmodes}), which are both found in the range from $10$ to $30$~cm.

\begin{figure}
	\centerline{\includegraphics[width=12cm]{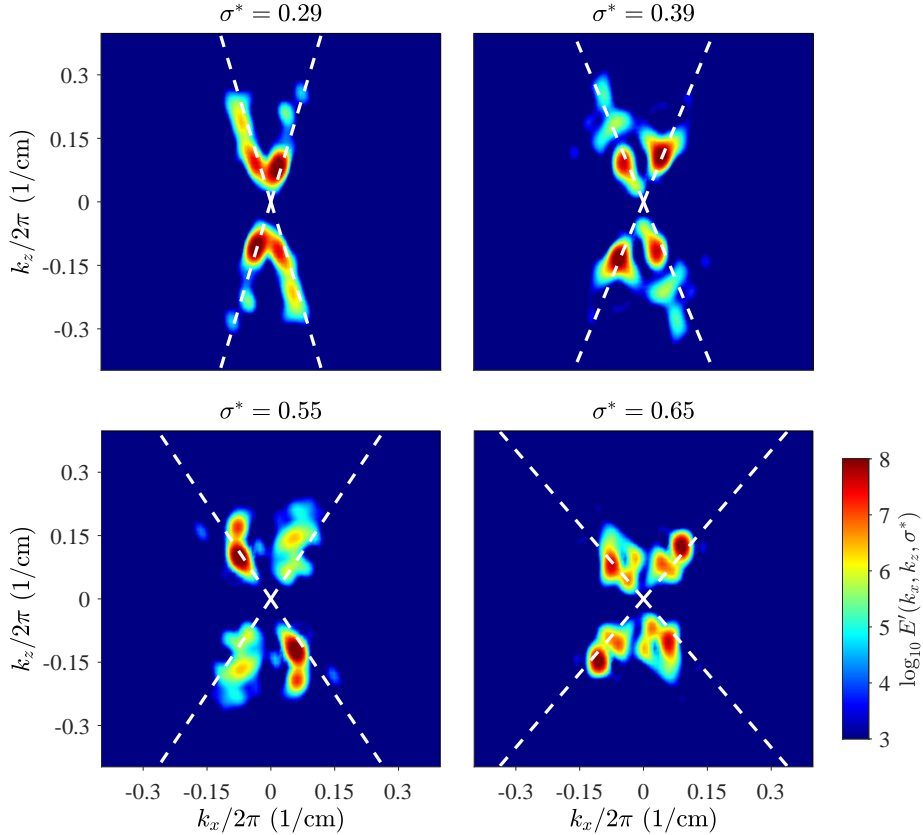}}
	\caption{Normalized spatio-temporal kinetic energy spectrum $E'(k_x,k_z,\sigma^*)$ for the experiment with tilted planes at $A=16$~mm for four values of the non-dimensional frequency, $\sigma^* = 0.29, 0.39, 0.55$ and $0.65$, corresponding to the velocity fields in Fig.~\ref{fig:champsig_tilted1}. These frequency values are highlighted in Fig.~\ref{fig_avec_bo} with blue dots. In each panel, the dashed lines represent the dispersion relation $|k_z|=|k_x|(1/\sigma^{*2}-1)^{1/2}$ of internal waves at frequency $\sigma^*$ and with $k_y=0$, i.e., propagating in the vertical measurement plane.}\label{fig:spk_spatiotempbov}
\end{figure}

Complementary to the velocity snapshots of Fig.~\ref{fig:champsig_tilted1}, we report in Fig.~\ref{fig:spk_spatiotempbov} the spatio-temporal kinetic energy spectrum $E'(k_x,k_z,\sigma^*)$ for the same experiment and for the same four frequencies $\sigma^*=0.29, 0.39, 0.55$, and $0.65$ (see Appendix~\ref{app:app2} for details on the calculation of $E'(k_x,k_z,\sigma^*)$). In each panel of Fig.~\ref{fig:spk_spatiotempbov}, the dashed lines represent the dispersion relation $|k_z|=|k_x|(1/\sigma^{*2}-1)^{1/2}$ of internal gravity waves that are invariant in the $y$-direction, i.e., with $k_y=0$. In the general case, plane internal waves verify the dispersion relation $|k_z|=(k_x^2+ k_y^2)^{1/2}(1/\sigma^{*2}-1)^{1/2}$ and will be associated in Fig.~\ref{fig:spk_spatiotempbov} to energy in the two angular sectors defined by $|k_z|\geq |k_x|(1/\sigma^{*2}-1)^{1/2}$. Nevertheless, we show in Appendix~\ref{app:appendix_spatiotemp2} that, even in the case of an ensemble of internal gravity waves with an axi-symetric distribution of wavevectors, we expect the spectrum $E'(k_x,k_z,\sigma^*)$ to be dominated by energetic spots close to the 2D dispersion relation $|k_z|=|k_x|(1/\sigma^{*2}-1)^{1/2}$, in a way similar to a flow composed only of waves propagating in the $(x,z)$ measurement plane ($k_y=0$).

In each panel of Fig.~\ref{fig:spk_spatiotempbov}, we see that most of the energetic regions are found close to the 2D dispersion relation ($k_y=0$), which observation is fully compatible with subharmonic flow modes composed of internal gravity waves. The energetic regions typically span a wave number range (in $k/2\pi$ units) from $0.05$ to $0.25$~cm$^{-1}$ which corresponds to lengthscales in the range from $4$ to $20$~cm. We also observe a slight tendency for the energetic regions to spread toward lower scales for decreasing frequencies in agreement with the observation of the velocity snapshots of Fig.~\ref{fig:champsig_tilted1}.

In conclusion, we have shown in this section that introducing slightly tilted planes at the top and at the bottom of our water tank prevents the emergence of eigenmodes of the fluid domain in the non-linear regime. Instead, when increasing the forcing amplitude we observe the progressive emergence of a continuum of energy over the whole subharmonic frequency range which is compatible with a statistically homogeneous ensemble of propagating internal gravity waves. For the experiment at the largest considered forcing amplitude $A=16$~mm, the mode at the forcing frequency corresponds to $70\%$ of the total flow kinetic energy and the subharmonic continuum of energy ---over the range $0.05\leq \sigma^*\leq 0.84$--- to about $17\%$. The rest of the energy is mainly associated to the energy peak at zero frequency ($8\%$ of the kinetic energy in the range $\sigma^*\leq 0.05$). This means that about $87\%$ of the total flow kinetic energy is carried by an ensemble of internal gravity waves continuously distributed in frequency over typically one decade.

\subsection{Low frequency mode}

\begin{figure}
	\centerline{\includegraphics[width=14cm]{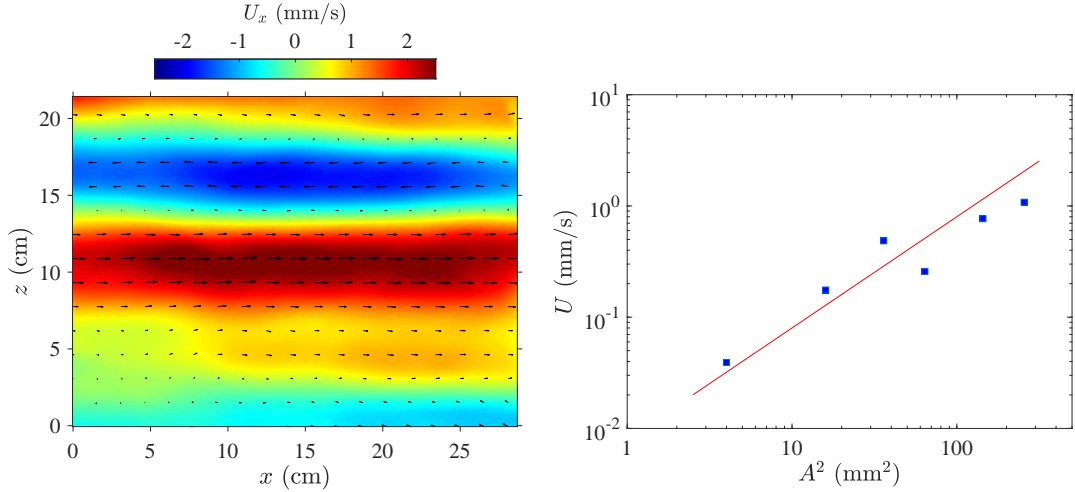}}
	\caption{(left) Velocity field ${\bf U} = \langle {\bf u} \rangle_t$ time-averaged over $200$ forcing periods in the statistically steady regime of the experiment at $A=16~$mm with the tilted planes. The color map shows the horizontal velocity component $U_x$. (right) Amplitude $U$ of the time-averaged velocity field as a function of the square of the forcing amplitude $A^2$ for the series of experiments with the tilted planes (in log scale). The straight line illustrates the scaling exponent $+1$. The amplitude $U$ is computed as $U=\sqrt{\langle {\bf U}^2\rangle_{\bf x}}$ where $\langle\, \rangle_{\bf x}$ stands for the spatial average over the PIV region.}\label{fig:champ_moyen}
\end{figure}

As we just noticed in the previous section, despite our flow is largely dominated by internal waves, the slow mode associated to the energy peak at zero frequency in the temporal energy spectra is starting to carry a significant part of the energy at the largest forcing amplitude. We therefore briefly characterize this mode in this section. In the left panel of Fig.~\ref{fig:champ_moyen}, we plot the measured velocity field time-averaged over $200$ forcing periods in the statistically steady regime of the experiment at $A=16~$mm. This field shows nearly horizontally invariant layers of horizontal velocity which are shearing each other in the vertical direction. Such a structure perfectly agrees with the expectations from the linearized Boussinesq equations at zero frequency $\sigma_0^*=0$ (and the internal wave dispersion relation~(\ref{eq:disp})) which predict horizontally invariant structures of horizontal velocity. The vertical thicknesses of the horizontal layers observed in Fig.~\ref{fig:champ_moyen}(left) are in the range from $5$ to $10$~cm.

The emergence of a slow flow mode that we observe here in an experiment where energy is injected in internal waves, with horizontally invariant layers of horizontal velocity, is reminiscent of the experiments of Rodda \textit{et al.}~\cite{Rodda2022} (presented in the introduction). Remarkably, Rodda \textit{et al.} also report the velocity field of their slow mode in two horizontal planes, which allow them to show that each horizontal layer observed in the vertical plane actually corresponds to a large eddy of vertical axis and with a size of the order of the size of the water tank. We cannot definitely confirm that the scenario is identical in our case. Nevertheless, it is very likely. Indeed, direct observations of the surface of our flow clearly confirm the presence of a large horizontal vortical circulation of the surface layer of the fluid at the size of the water tank.

As highlighted by Rodda \textit{et al.}~\cite{Rodda2022}, the slow mode we observe in our experiments forced by internal waves is reminiscent of the vertically sheared horizontal flow (VSHF) which emerges in direct numerical simulations of strongly stratified turbulence~\cite{Smith2002,Waite2006,Augier2015,Maffioli2020,Lam2020,Lam2021}. In these homogeneous turbulence simulations, the VSHF, which results from inverse energy transfers from the injection scale, is allowed to be strictly horizontally invariant thanks to the periodic boundary conditions. Obviously, this is not the case in experiments in a water tank where the most horizontally invariant mode that can emerge is the large-scale vortical mode we observe.

The mode at zero frequency of our experiments can also result from non-linear 
processes directly affecting the periodic flow at the forcing frequency. Such 
processes have indeed often been reported in periodically forced stratified 
fluid experiments or numerical simulations where they may take place either in 
the bulk of the flow~\cite{Sutherland2006,Bordes2012b,Jamin2021} or in the 
vicinity of the wavemakers~\cite{Ermanyuk2008,King2009,Fan2020}. In some of 
these works, the non-linear process has moreover been identified to be of 
streaming type~\cite{Bordes2012b,Fan2020,Jamin2021} or of Stokes drift 
type~\cite{Sutherland2006}. Steady-streaming and Stokes drift take place in 
periodic flow with a spatially inhomogeneous amplitude and result in the 
production of a steady flow proportional to the square of the base flow 
velocity~\cite{Sutherland2006,Bordes2012b,Fan2018}. 

To test this scaling, we plot in the right panel of Fig.~\ref{fig:champ_moyen}, 
the spatial rms velocity of the time-averaged velocity field as a function of 
the square $A^2$ of the amplitude of the cylinders oscillations in our 
experiments with tilted planes. Fig.~\ref{fig:champ_moyen}(right) reveals a 
behavior compatible with a linear scaling law and therefore with a quadratic 
non-linear process, such as steady streaming or Stokes drift, affecting the 
base flow produced by the oscillating cylinders. Nevertheless, a more thorough 
experimental and theoretical study is necessary to draw clear conclusions on 
the non-linear mechanism at the origin of the mean flow in our experiments, 
which study is out of the scope of this article.

\subsection{Kinetic energy spatial spectra}

\begin{figure}
	\centerline{\includegraphics[width=12cm]{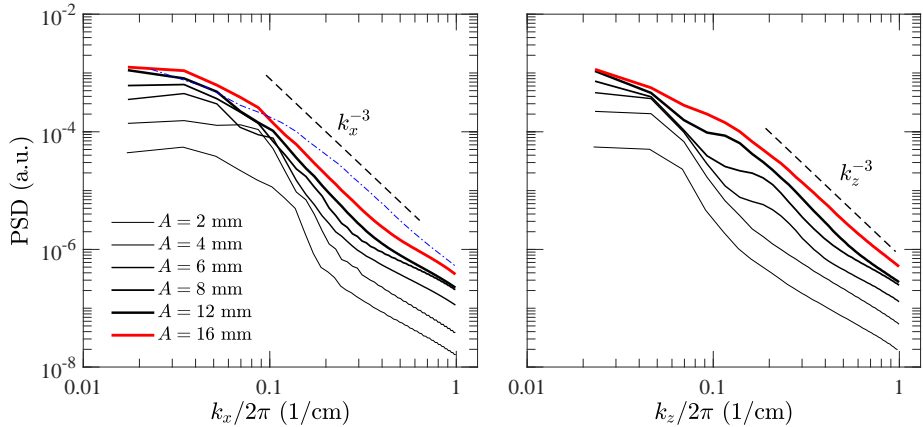}}
	\caption{One dimensional spatial kinetic energy spectra as a function of 
	the horizontal wavenumber $k_x$ (left) and the vertical wavenumber $k_z$ 
	(right) for the experiments with the tilted planes at different forcing 
	amplitudes, $A=2, 4, 6, 8, 12$ and $16$~mm from bottom to top curves.
To quantify the flow anisotropy, we also report in the left panel, as a 
thin blue dashed-dotted curve, the $k_z$-spectrum for the 
experiment at $A=16$~mm (i.e., the same spectrum as the one in red thick line in the right 
panel).}\label{fig:spatial_spk}
\end{figure}

Having retrieved hints of the premises of a wave-turbulence like regime in our series of experiments with the tilted planes, it is interesting to look in Fig.~\ref{fig:spatial_spk} at the spatial power spectral densities of the measured velocity field for these experiments (Tab.~\ref{tab:freqb2}). All these spectra are one dimensional, either as a function of the horizontal wavenumber $k_x$ or of the vertical wavenumber $k_z$. This means one recovers the average kinetic energy of the (measured) velocity field when these 1D spectra are integrated over $k_x$ or $k_z$, respectively. We estimate these spectra by computing the 1D spatial Fourier transform of the instantaneous two-point velocity correlation, along $x$ or $z$, using the Wiener-Khinchin theorem before taking the temporal and spatial average (over the remaining spatial direction).

In the left panel of Fig.~\ref{fig:spatial_spk}, we show the 1D spatial kinetic energy spectra as a function of the horizontal wavenumber $k_x$ for the different values of the forcing amplitude. For the linear experiment at the lowest amplitude, $A = 2$~mm, the energy of the flow is gathered in horizontal wavenumbers $k_x/2\pi$ below $0.13$~cm$^{-1}$ and the spectrum rapidly decays at larger wavenumbers. These energetic wavenumber values correspond to horizontal wavelengths of $8$~cm and larger, in good agreement with the observation of the forced mode suggesting horizontal wavelengths in the range $15\pm 5$~cm. As the forcing amplitude is increased, we observe the progressive increase of the energy content at horizontal wavenumber $k_x/2\pi$ larger than $0.13$~cm$^{-1}$, i.e. at smaller horizontal scales. At the largest forcing amplitude $A = 16$~mm, one could argue that a power-law behavior, compatible with a decay exponent $-3$, has emerged over the wavenumber range $0.09$~cm$^{-1}\leq k_x/2\pi\leq 0.45$~cm$^{-1}$.

The corresponding spatial energy spectra as a function of the vertical wavenumber $k_z$ are reported in the right panel of Fig.~\ref{fig:spatial_spk}. At the lowest forcing amplitude $A = 2$~mm, i.e. in the linear regime, we observe energy below the cut-off wavenumber $k_z/2\pi \simeq 0.07$~cm, which corresponds to vertical wavelengths larger than $14$~cm. The direct observations of the velocity snapshots of the mode at the forcing frequency $\sigma_0^*=0.94$ (as seen in Fig.~\ref{fig:champ4mm}) evidence vertical wavelengths in the range $25\pm 5$~cm which is roughly consistent with the energy spectrum observed here for $A=2$~mm. We should however highlight that the vertical size of our velocity measurement field, of $22$~cm, is of the same order as the vertical scales of the forced mode (and even generally smaller). It is therefore clear that the spatial spectrum as a function of $k_z$ cannot accurately account for the largest energetic vertical scales of the flow associated the forced mode. Nevertheless, as for the horizontal energy spectra, when increasing the forcing amplitude, we observe the progressive emergence of an energy continuum at larger (properly-resolved) wavenumbers $k_z$, i.e. at smaller vertical scales. For the experiment at the largest forcing amplitude $A = 16$~mm, we observe over the wavenumber range $0.20$~cm$^{-1}\leq k_x/2\pi\leq 0.90$~cm$^{-1}$ (again of about half-a-decade) a behavior roughly compatible with a power law of exponent $-3$.

\section{Conclusion}

In this article, we present laboratory experiments conducted in a linearly stratified fluid forced by a set of $24$~wavemakers. Our forcing device injects energy in an ensemble of internal gravity wave beams at frequency $\sigma_0=0.94\,N$ (where $N$ is the buoyancy frequency) which approaches statistical homogeneity and axi-symmetry. When the forcing amplitude is increased, the flow non-linearities emerge through the establishment of a set of several internal wave modes at discrete subharmonic frequencies (i.e. lower than $\sigma_0$). These modes, the most energetic of which are shown to be resonant eigenmodes of the fluid domain, are in temporal triadic resonances with the forced waves and between themselves. This scenario is reminiscent of the one observed in several recent experiments where a turbulent flow is also forced through internal waves in a stratified fluid~\cite{Brouzet2017,Savaro2020,Davis2020}. The discretization of the energy in frequency and in wavenumber that we observe prevents the turbulent flow from approaching the ``wave turbulence'' regime described in the Weak/Wave Turbulence Theories~\cite{McComas1981,Caillol2000,Lvov2004,Lvov2010,Dematteis2021}. In these theoretical descriptions, the turbulent cascade is indeed carried by a statistical ensemble of weakly non-linear and propagating waves in an infinite domain which forms an energy continuum in the frequency and wavenumber spaces. 

Nevertheless, we identify a slight but crucial modification of our experimental setup which allows to efficiently inhibit the emergence of the wave eigenmodes. It consists in introducing two slightly tilted panels in the fluid domain, one at the top and one at the bottom. The change in wavelength induced by these inclined planes when the waves are reflecting on them is small to moderate for most frequencies in the internal wave range ($\sigma \leq N$) but sufficient to prevent the formation of standing modes in the experimental cavity. This modified setup finally allows, in the non-linear regime, the development of a continuum of energy over one decade in the wave frequency range together with a disappearance of the energy peaks associated to eigenmodes of the fluid domain. Moreover, we show that this energy continuum is mainly carried by internal gravity waves verifying the wave dispersion relation.

Eventually, we achieve with our modified setup a turbulent flow approaching a three-dimensional internal wave turbulence regime in which no discretization of the energy in frequency and wavenumber is observed. In this configuration, our 1D spatial energy spectra, as a function of the horizontal and vertical wavenumbers, both tend to draw a power law with an exponent $-3$ at the largest explored Reynolds number, which result is compatible with the recent observations of Le Reun \textit{et al.}~\cite{LeReun2018} and Davis \textit{et al.}~\cite{Davis2020}. However, these power-law behaviors still remain questionable since they are observed over no more than half-a-decade of wavenumbers, which was also the case in Refs.~\cite{LeReun2018,Davis2020}. Besides, it is possible that the regime we observe at the largest forcing Reynolds number in our experiments, when power-law energy spectra seem to emerge, might reveal a system of internal waves at the onset of strong non-linearities.

Weak Wave Turbulence theory applied to internal waves in stratified fluids has long been proposed as a possible candidate to explain the oceanic energy spectrum at small scales~\cite{McComas1981,Caillol2000,Lvov2004,Lvov2010,Dematteis2021}. Nevertheless, the development of this analytical formalism is still a matter of debate and providing data of experiments in which a wave turbulence regime with ``fully developed'' energy cascade is observed seems important to move forward. In this context, it is clear that experiments at larger Reynolds numbers shall be conducted in order for the energy cascade to develop over a larger range of spatial scales than what we achieved in this work. This increase in Reynolds number cannot be achieved by increasing the typical velocity of the forcing since it would be accompanied by an increase in the forcing Froude number and lead to a strongly non-linear turbulence and to a strong mixing of the stratification in density.
The most promising (and probably the only) experimental track to achieve a genuine weakly non-linear internal wave turbulence regime in the lab, with spatial energy spectra with ``well developed'' power laws, is to significantly upscale the size of the experiment and in particular of the injection wavelength, while still inhibiting finite size effects and the emergence of fluid domain eigenmodes.

\begin{acknowledgments}
This work was supported by a grant from the Simons Foundation (651461, PPC). We thank B. Gallet for stimulating scientific discussions. We acknowledge J. Amarni, A. Aubertin, L. Auffray, C. Manquest, R. Pidoux and B. Semin for experimental help. 
\end{acknowledgments}

\appendix

\section{Computation of the spatio-temporal spectra}
\label{app:app2}

The normalized spatio-temporal spectra $E'(k_x,k_z,\sigma)$ presented in the article are computed from the two-dimensional two-component measured velocity field ${\bf u}({\bf x},t)=(u_x({\bf x},t),u_z({\bf x},t))$ in the following way (${\bf x}=(x,z)$ is the position vector in the vertical measurement plane in cartesian coordinates).

We first compute the sum $R({\bf r},\sigma)$ of the two-point correlations in the measurement plane of the temporal Fourier transform $\tilde{u}_i({\bf x},\sigma)$ of the two components of the measured velocity field $u_i({\bf x},t)$ defined by
\begin{equation}
   R({\bf r},\sigma)=\langle \tilde{u}_i({\bf x},\sigma)\tilde{u}_i^\star({\bf x}+{\bf r},\sigma) + \tilde{u}_i^\star({\bf x},\sigma)\tilde{u}_i({\bf x}+{\bf r},\sigma) \rangle_{\bf x}.
\end{equation}
In this expression, the star denotes the complex conjugate, the brackets $\langle \, \rangle_{\bf x}$ stand for the spatial average over the measurement area and $i$ refers to a sum over $i=x$ and $z$. 

The two-point correlation $R({\bf r},\sigma)$ is defined for separation vectors ${\bf r}=(r_x,r_z)$ in the range $-L_x \leq r_x \leq L_x$ and $-L_z \leq r_z \leq L_z$ where $L_x$ and $L_z$ are the sizes of the measured velocity field in the horizontal and vertical directions, respectively. The two-point correlation $R({\bf r},\sigma)$ is then multiplied by a Hann function in the vertical and horizontal directions (of respective width $2\, L_x$ and $2\, L_z$) before it is finally extended by zero values to the range $-1.5 \, L_x \leq r_x \leq 1.5 \,L_x$ and $-1.5 \,L_z \leq r_z \leq 1.5 \,L_z$ (zero padding method).

We finally compute the two-dimensional spatial Fourier transform of the obtained function of ${\bf r}=(r_x,r_z)$ (and frequency $\sigma$) and normalize it by its integral over the wavevector space $(k_x,k_z)$ for each frequency $\sigma$ in order to get the normalized spatio-temporal spectra $E'(k_x,k_z,\sigma)$.

\section{Spatio-temporal kinetic energy spectrum for a statistically axisymmetric distribution of internal gravity waves}\label{app:appendix_spatiotemp2}

In Fig.~\ref{fig:spk_spatiotempbov}, we report the spatio-temporal energy spectrum $E(k_x,k_y,\sigma)$ of the measured velocity field for the experiment with the tilted planes at $A=16$~mm for four values of the frequency $\sigma$. Since we measure experimentally only the two components $u_x$ and $u_z$ of the velocity field in the vertical plane ($x,y=y_0,z$), the interpretation of this spatio-temporal spectrum is not straightforward. In order to help with this interpretation, we compute in this appendix the analytical expression of $E(k_x,k_y,\sigma)$ for a flow made of an ensemble of independent plane internal waves with an axisymmetric wavevector statistics.

First, we introduce the 3-dimensional (3D) spatio-temporal Fourier transform 
\begin{equation}
a_i(k_x,y,k_z,\sigma)=\frac{1}{(2\pi)^{3/2}} \iiint u_i(x,y,z,t) e^{-i(k_x x +k_z z - \sigma t)}\,dx\,dz\,dt
\end{equation}
and the 4-dimensional (4D) spatio-temporal Fourier transform 
\begin{equation}
b_i(k_x,k_y,k_z,\sigma)=\frac{1}{(2\pi)^{1/2}} \int a_i(k_x,y,k_z,\sigma) e^{-i k_y y }\,dy
\end{equation}
of the component $u_i$ of the velocity field ${\bf u}({\bf x},t)$ where ${\bf x}=(x,y,z)$ is the position in Cartesian coordinates, ${\bf k}=(k_x,k_y,k_x)$ the corresponding wavevector and $i=x,y$ or $z$. 
Ignoring the component of the velocity along the direction $y$ as done in the data analysis, the 3D spatio-temporal power spectral density of the velocity field is defined by
\begin{equation}
S_{3D}(k_x,y,k_z,\sigma)= \frac{a_x\,a_x^\star+a_z\,a_z^\star}{\tau L_x L_z}\, ,
\end{equation}
and the 4D one by 
\begin{equation}\label{eq:S4D}
S_{4D}(k_x,k_y,k_z,\sigma)= \frac{b_x\,b_x^\star+b_z\,b_z^\star}{\tau L_x L_y L_z}\, ,
\end{equation}
where the stars denote the complex conjugate, $\tau$ the duration and $L_x$, $L_y$ and $L_z$ the distances (in directions $x$, $y$ and $z$, respectively) over which the Fourier transforms have been computed. In Fig.~\ref{fig:spk_spatiotempbov}, we report the 3D spatio-temporal energy spectrum $E(k_x,k_y,\sigma)=S_{3D}(k_x,y=y_0,k_z,\sigma)$ evaluated in the specific plane $y=y_0$.

Let us now consider a plane internal gravity wave of non-dimensional frequency $\sigma^*=\sigma/N$ and wavevector ${\bf k}=(k_x,k_y,k_z)$. According to the internal gravity wave dispersion relation~(\ref{eq:disp}), the horizontal $k_\perp=\sqrt{k_x^2+k_y^2}$ and vertical $k_\parallel=|k_z|$ wavenumbers of the plane wave verify $k_\perp = k \sigma^*$ and $k_\parallel = k \sqrt{1-\sigma^{*2}}$, where $k=|{\bf k}|$ is the norm of the wavevector. Then, because of incompressibility (${\bf u} \cdot {\bf k}=0$), the velocity field of this plane wave can be written
\begin{eqnarray}
u_x & = & u_0 \sqrt{1-\sigma^{*2}} \frac{k_x}{k_\perp} \cos{({\bf k}\cdot{\bf x} - \sigma t +\varphi)}\, ,\label{eq:ux0}\\
u_y & = & u_0 \sqrt{1-\sigma^{*2}} \frac{k_y}{k_\perp} \, \cos{({\bf k}\cdot{\bf x} - \sigma t +\varphi)} \, ,\label{eq:uy0}\\
u_z & = & - u_0 \sigma^*\, \cos{({\bf k}\cdot{\bf x} - \sigma t +\varphi)}\, .\label{eq:uz0}
\end{eqnarray}

Let us then consider an ensemble of independent plane internal gravity waves, all with the same amplitude $u_0$, the same non-dimensional frequency $\sigma_0^*$ and the same wavenumber $k_0=|{\bf k}|$, but with an axisymmetric wavevector statistics, meaning that the azimuthal angle $\tan^{-1}(k_y/k_x)$ is uniformly distributed between $0$ and $2\pi$. According to Eqs.~(\ref{eq:ux0}-\ref{eq:uz0}) and considering only the $u_x$ and $u_z$ components of the velocity field, the 4D spatio-temporal power spectral density (defined in Eq.~\ref{eq:S4D}) for such an ensemble of plane waves writes
\begin{eqnarray}
S_{4D}(k_x,k_y,k_z,\sigma) &\propto& u_0^2 \left[(1-\sigma_0^{*2}) \frac{k_x^2}{k_{\perp, 0}^2}+ \sigma_0^{*2}\right] \nonumber \\
&&\times \delta(\sigma \pm \sigma_0)\delta(k_x^2+k_y^2-k_{\perp,0}^{2})\delta(k_z \pm k_{\parallel,0})\, ,\label{eq:S4Dbis}
\end{eqnarray}
where $\delta(x \pm a) = \delta(x-a) + \delta(x +a)$ and $k_{\perp,0} = k_0 \sigma_0^*$ and $k_{\parallel,0} = k_0 \sqrt{1-\sigma_0^{* 2}}$ are the horizontal and vertical wavenumbers of all the waves involved in the statistics, respectively.

Using the statistical homogeneity of the considered velocity field, the spectrum $E(k_x,k_z,\sigma)$ is related to the 4D spectrum through the Parseval theorem:
\begin{eqnarray}
E(k_x,k_z,\sigma) =  \frac{1}{L_y}\int S_{3D}(k_x,y,k_z,\sigma)\, dy = \int S_{4D}(k_x,k_y,k_z,\sigma)\, dk_y\,.
\end{eqnarray}
The integration of the delta function $\delta(k_x^2+k_y^2-k_{\perp, 0}^2)$ in~(\ref{eq:S4Dbis}) finally yields
\begin{eqnarray}
E(k_x,k_z,\sigma) \propto \, u_0^2 \left[(1-\sigma_0^{* 2}) \frac{k_x^2}{k_{\perp, 0}^2}+ \sigma_0^{* 2}\right]\frac{1}{\sqrt{k_{\perp, 0}^2-k_x^2}} \delta(\sigma \pm \sigma_0)\delta(k_z \pm k_{\parallel,0})\, , \label{eq:spec_spatiotemp}
\end{eqnarray}
when $|k_x|\leq k_{\perp, 0}$ and $E(k_x,k_z,\sigma)=0$ otherwise.

In the expression~(\ref{eq:spec_spatiotemp}) of the spectrum $E(k_x,k_z,\sigma)$, we see that the term $1/\sqrt{k_{\perp, 0}^2-k_x^2}$, stemming from the ``projection'' of the 4D spectrum on the ($k_x,k_z$) plane, diverges when $|k_x|$ tends towards $k_{\perp, 0}$ (from low values, i.e. for $|k_x|\leq k_{\perp, 0}$). Since $k_{\perp, 0}$ is simply the value taken by $|k_x|$ for waves at frequency $\sigma_0$ and propagating in the measurement plane, the divergence of $E(k_x,k_z,\sigma)$ suggests that an axisymmetric distribution of internal gravity waves will have, with $E(k_x,k_z,\sigma)$, a spectral signature resembling the one of waves propagating in the measurement plane. Besides, the term $[(1-\sigma_0^{* 2}) k_x^2/k_{\perp, 0}^2+ \sigma_0^{* 2}]$, stemming from the fact we do not consider the $u_y$ component of the velocity, strengthens this diverging behavior, yielding an additional increase of the energy density as $|k_x|$ increases from zero towards the limit $|k_x|=k_{\perp, 0}$. One can note that this additional effect is stronger as the wave non-dimensional frequency decreases from $1$ to $0$.

\end{document}